\documentclass[
reprint,
onecolumn,
nofootinbib,
amsmath,amssymb, aps
]{revtex4-2}
\pdfoutput=1

\usepackage[utf8]{inputenc}
\usepackage{graphicx}
\usepackage{amsmath}
\usepackage{amssymb}
\usepackage{xcolor}

\begin{document}

\title{Unified description of corpuscular and fuzzy bosonic dark matter II: Dissipation and stochastic forces}
\author{Nick P. Proukakis$^{1}$}
\email[E-mail: ]{nikolaos.proukakis@newcastle.ac.uk}
\author{Gerasimos Rigopoulos$^{1}$}
\email[E-mail: ]{gerasimos.rigopoulos@newcastle.ac.uk}
\author{Alex Soto$^{1}$}
\email[E-mail: ]{arsoto1@uc.cl}
\affiliation{$^{1}$ School of Mathematics, Statistics and Physics, Newcastle University, Newcastle upon Tyne, NE1 7RU, UK}
\date{January 2024}

\begin{abstract}
{\noindent We extend our previous work (Proukakis {\em et al.}, Phys.~Rev.~D~108,~083513 (2023)~\cite{Proukakis:2023nmm}) on the dynamics of bosonic, non-relativistic and  self-interacting dark matter that  simultaneously contains both a ``fuzzy'' low-momentum component and one with higher momenta that may be well approximated as a collection of distinct particles and described by a corresponding phase-space distribution. Starting from the non-relativistic Schwinger-Keldysh action and working beyond leading-order in the Keldysh basis fields, encoding stochastic fluctuations of the slow modes and all fluctuations of the fast modes, we obtain stochastic self-consistently coupled Gross-Pitaevskii, collisional Boltzmann kinetic and Poisson equations.
Our final set of equations, which feature various collisional (dissipative and scattering) contributions and two corresponding independent stochastic force terms, are consistent with generalized fluctuation-dissipation type relations in the limit of thermal equilibrium between the particles.
}  
\end{abstract}

\maketitle

\section{Introduction}

Over the last decade or so an alternative idea for the microphysics of dark matter has been rising in popularity  among a growing number of cosmologists - see \cite{2016PhR...643....1M, 2021ARA&A..59..247H, 2021A&ARv..29....7F, OHare:2024nmr,Matos:2023usa} for recent reviews and references to the literature. Already introduced more than 2 decades ago \cite{Hu:2000ke} and often called fuzzy dark matter (FDM) or $\psi$DM, the model assumes that dark matter is composed of ultra-light bosons that can exhibit wave phenomena on galactic scales (indicatively $\lesssim$ 1 kpc). Interestingly, in addition to reproducing the successful predictions of  the corpuscular model of Cold Dark Matter (CDM) at early times and on larger scales, such a model -- and its extension to include self-interactions -- does in fact present distinct phenomenology on short scales~\cite{Boehmer:2007um,Chavanis_2011,PhysRevD.84.043532,Rindler-Daller:2011afd,PhysRevD.89.083536,2014NatPh..10..496S,2016PhRvD..94d3513S,PhysRevD.96.063505,2017MNRAS.471.4559M,PhysRevD.97.023529,Veltmaat:2018dfz,Chavanis:2020jkc,Chavanis:2020rdo,May_2021,Delgado:2022vnt,Kirkpatrick:2021wwz,Chakrabarti:2022owq,Hartman:2021upg,Hartman:2022cfh,Nori:2022afw,Dave:2023wjq,Mocz:2023adf,Indjin:2023hno,Stallovits:2024cpg,Painter:2024rnc,Yang:2024hvb}.
Mathematically, this model is normally described by the Schr\"{o}dinger-Poisson or Gross-Pitaevskii-Poisson systems of coupled equations, essentially a wave equation that includes the (Newtonian) self-Gravity of the classical field representing the bosons. 
However, matter made up of bosonic particles can fall anywhere between the limiting descriptions of a completely continuous field (wavelike) state and a particle-like one, depending on the underlying velocity distribution. Even if a wavelike description is ultimately justified for all modes due to high phase space occupancy $\frac{\rho}{m} \lambda_\mathrm{dB}^3=\frac{h^3}{m^4} \frac{\langle\rho\rangle}{\langle|{v}|\rangle^3} \gtrsim \mathcal{O}(1)$, a situation expected in cosmologically relevant configurations if the bosons are light enough,  it may nonetheless be convenient or desirable to describe a sub-population of the particles using an N-body phase space description. For example this would be the case if there is a large range of velocities and therefore the related de Broglie wavelengths span a correspondingly large range without a single peak around a dominating characteristic value - see \cite{Proukakis:2023txk} for an application of such a configuration in the case of cosmological perturbation theory, and similarly \cite{Schwabe:2020eac,Lague:2023wes} for situations where a mixed configuration has been studied. For a coherence loss mechanism in the early universe see \cite{Brandenberger:2023idg}.           

In our earlier work \cite{Proukakis:2023nmm}, to which we also refer the reader for a more extended discussion of the context of these computations and the relation to other known formalisms~\cite{2016PhR...643....1M,2014NatPh..10..496S, 2016PhRvD..94d3513S, 2017MNRAS.471.4559M, Veltmaat:2018dfz, May_2021, Nori:2022afw,Boehmer:2007um, Chavanis_2011,PhysRevD.84.043532,Rindler-Daller:2011afd, Guth:2014hsa, 2014NatPh..10..496S, Schwabe:2020eac, Chavanis:2020jkc, Chen:2021oot, Kirkpatrick:2021wwz, Hartman:2022cfh, Delgado:2022vnt, Mocz:2023adf, Lague:2023wes, Yang:2024hvb, Stallovits:2024cpg, Painter:2024rnc, Proukakis:2023txk, Bernardeau+, Angulo+Hahn,zaremba1999dynamics,griffin_nikuni_zaremba_2009,proukakis2008finite,Lee_2016-ZNG-review}, we derived a set of coupled equations that describe the slow modes via a Gross-Pitaevskii equation and the fast modes via a phase space kinetic equation, also including their mutual gravitational attraction. In this work we extend our earlier computations to include fluctuations of the fast modes and obtain equations that go beyond the leading semi-classical dynamics, finding a novel set of two `dissipative/collisional' kernels and two associated stochastic noise forces, one additive and one multiplicative, which are related to the fluctuations of the high momentum particles. The respective noises and dissipative/collisional kernels come in pairs, each exhibiting the expected structure in terms of the fast modes' phase space distribution. 
We further comment on the nature of the fast/slow mode split and the corresponding subtle distinction between ``coherent'' and ``incoherent'' modes of the system, and how that relates to their description in terms of coupled wave and kinetic equations.
We do not assume thermal equilibrium and the distribution of the higher momentum particles is dynamically evolved via a collisional Boltzmann equation. However, if a thermal Bose-Einstein distribution is assumed for the fast modes, fluctuation-dissipation-like relations between the stochastic noises and the dissipative/collisional kernels are retrieved.  

The next section presents the derived stochastic-dissipative dynamical equations, while section \ref{Steps} describes the steps necessary for their derivation, with various intermediate technical computations relegated to Appendices. Section \ref{Thermal-plus} briefly presents 
the emergence of fluctuation-dissipation relations in the thermal limit, with some final thoughts and pointers to possible future work presented in section \ref{Final}.

\section{Stochastic - Dissipative dynamical equations for coherent and incoherent modes}\label{Equations}

We start by considering the relativistic covariant action for an ultralight bosonic particle coupled minimally to gravity, given by
\begin{equation}
S = \int d^4 x \sqrt{- g} \bigg( \frac{1}{16 \pi G} R - \frac{1}{2} {g}^{\mu
\nu} \partial_{\mu} \phi \partial_{\nu} \phi - m^2 \phi^2 - \frac{m^2\lambda}{2}\phi^4 \bigg) \,,
\end{equation}
where $G$ is the gravitational constant and $R$ is the Ricci scalar containing the spacetime metric ${g}_{\mu\nu}$ and in consequence the system's gravity. The quantity ${g}$ that appears in the action is the determinant of this metric. We have used the shorthand $d^4 x=dt d^3 \mathbf{r}$ which we will also employ throughout the paper and we work in natural units ($c=\hbar=1$). The scalar field $\phi$ represents a bosonic particle of mass $m$ and we consider a self-interaction with coupling $\lambda$. If we assume a flat spacetime metric with a small scalar perturbation describing weak gravity in the Newtonian gauge, we have
\begin{equation}
d s^2 = - (1 + 2 V) d t^2 + (1 - 2 V) \delta_{i j} d x^i d x^j \;,
\end{equation}
where $V \ll 1$ will describe the Newtonian gravitational potential in the non-relativistic limit. Considering the action up to second order in $V$, redefining $\lambda=\frac{g}{m^2}$ \cite{Suarez:2015uva}, writing 
\begin{equation}
\phi = \frac{1}{\sqrt{2 m}} \left( \psi e^{- i m t} +
\psi^{\ast} e^{i m t} \right) \;,
\end{equation}
and going to the non-relativistic limit, we obtain
\begin{equation}
\label{actionfdm}
S = \int d t d^3 \mathbf{r} \left( i \psi^{\ast} \dot{\psi} + \frac{1}{2
m} \psi^{\ast} \nabla^2 \psi - \frac{g}{2} | \psi |^4 - m V | \psi |^2 +
\frac{1}{8 \pi G} V\nabla^2 V \right) \;,
\end{equation}
which will be our starting point action functional for the subsequent analysis. 

For studying cosmic large scale structures it would be desirable to also consider the expansion of the Universe. In such case, we can start from a FLRW metric with a scalar perturbation $V$ in the Newtonian gauge, defined by
\begin{equation}
d s^2 = - (1 + 2 V) d t^2 + a^2(1 - 2 V) \delta_{i j} d x^i d x^j \;,
\end{equation}
where $a(t)$ is the scale factor of the universe. Following the same steps as before we can arrive at the action
\begin{equation}
S = \int d t d^3 \mathbf{r} \ a^3 \bigg( i \psi^{\ast} \dot{\psi} + \frac{3}{2} i H | \psi |^2 + \frac{1}{2 m a^2} \psi^{\ast} \nabla^2 \psi -
\frac{g}{2} | \psi |^4 - m V | \psi |^2 - \frac{1}{8 \pi G a^2} (\nabla V)^2 \bigg) \;.
\end{equation}
where now $\mathbf{r}$ refers to comoving coordinates and we have integrated by parts in the last term, ignoring boundary terms. Here we note the presence of an extra term, $ \frac{3}{2} i H | \psi |^2$, where  $H=\dot{a}/a$ is the Hubble parameter. 
It is convenient to make a redefinition to the \emph{comoving field} $\psi\to a^{-3/2}\psi$ and thus obtain
\begin{equation}
\label{actionexp}
S = \int d t d^3 \mathbf{r} \left( i \psi^{\ast} \dot{\psi} + \frac{1}{2 m
a^2} \psi^{\ast} \nabla^2 \psi - \frac{g}{2 a^3} | \psi |^4 - m V | \psi |^2 -
\frac{a}{8 \pi G} (\nabla V)^2 \right) \;,
\end{equation}
which is the same action as that  used in our previous work~\cite{Proukakis:2023nmm}. In order to be able to apply the formalism developed in the present  paper it is important to remark that we must require a regime where the scale factor $a$ varies sufficiently slowly. Furthermore, from now on we will add to action \eqref{actionexp} an external potential $V_{ext}$
\begin{equation}
\label{eqini}
S = \int d t d^3 \mathbf{r}  \left( i \psi^{\ast} \dot{\psi} + \frac{1}{2
m a^2} \psi^{\ast} \nabla^2 \psi - \frac{g}{2a^3} (\psi^{\ast}\psi)^2 - V_{ext} \psi^{\ast}\psi - m V \psi^{\ast}\psi + \frac{a}{8\pi G}
V\nabla^2 V  \right)\:.
\end{equation}
While such an external potential is usually zero in typical cosmological contexts we nonetheless include it here, as it is possible to add external point particle potentials such as those stemming from black holes residing in the centres of halos \cite{Davies:2019wgi}. Note that the parameter $m$ attached to the second-to-last term now plays the role of the strength of the coupling of the boson with the field $V$. We will consider this parameter to be small, in the same way as we will consider the parameter $g$ to be small, since we will work with both parameters in a perturbative treatment.

Note that by using a Hubbard-Stratonovich transformation in the term involving the gravitational interaction, the non-relativistic action \eqref{eqini} is equivalent to 
\begin{equation}
S = \int d t \bigg(i \int d^3 \mathbf{r} \ \psi^{\ast} \dot{\psi}- H \bigg) \;,
\end{equation}
where the Hamiltonian $H$ has the following form
\begin{equation}
\label{hamiltoniangen}
H = \int d^3 \mathbf{r} \left( - \frac{1}{2 ma^2} \psi^{\ast} \nabla^2 \psi +
V_{ext}\psi^\ast \psi \right) + \frac{1}{2} \int d^3 \mathbf{r} \int d^3 \mathbf{r}'  \psi^\ast(t,\mathbf{r}) \psi^\ast(t,\mathbf{r}') U (\mathbf{r}, \mathbf{r}') \psi(t,\mathbf{r}')\psi(t,\mathbf{r}) \;.
\end{equation}
%
The term $U(\mathbf{r},\mathbf{r}')$ describes the bosonic interaction potential, which can be written in terms of two parts, in the form
\begin{equation}
\label{interaction}
U (\mathbf{r}, \mathbf{r}') = g \frac{\delta (\mathbf{r} - \mathbf{r}')}{a^3} -\frac{G m^2}{a|\mathbf{r}- \mathbf{r}'|} \;,
\end{equation}
where the second term denotes the usual gravitational potential. The first term denotes a contact interaction, stemming from the $\phi^4$ self-interaction term 
and parameterized by the coupling constant $g$. The incorporation of such an interaction term is very well motivated. For example, attractive self-interactions appear in theoretical considerations for light CP-odd axions \cite{PhysRevLett.40.223,PhysRevLett.40.279,Peccei:1977hh,PhysRevD.16.1791}, while particle models with repulsive self-interactions have been also constructed \cite{Fan:2016rda}. The interest in self-interacting models has increased and some constraints have been placed on the coupling \cite{PhysRevD.89.083536,PhysRevD.96.063505,Chakrabarti:2022owq,Hartman:2021upg,Delgado:2022vnt,PhysRevD.97.023529}.

For the convenience of the readers not interested in following all the technical steps, we present below the equations that consitute the main results of this paper.  The steps required to obtain them are explained in the next section with further technical details relegated to the appendices. Working in the Schwinger-Keldysh formalism with action \eqref{eqini}, we split the field into slow ($\Phi_0$) and fast ($\varphi$) components 
\begin{equation}
\psi = \Phi_0 +
\varphi \;,
\end{equation}
and derive a phase space density $f(\mathbf{r},\mathbf{p},t)$ from the Wigner transform of the Keldysh component of the two point correlator $G^K(x,x')=\langle \varphi^{\ast} (x) \varphi (x') \rangle$. We will also be referring to $\Phi_0$ and $\varphi$ as the ``coherent" and ``incoherent", or ``particle", parts respectively\footnote{As discussed in section~\ref{sec:cut} the ``coherent'' part refers to both the pure ``condensate'' mode and a subset of modes adjacent to it which maintain some partial spatio-temporal coherence.}. Their defining difference is that $\Phi_0$ and $\varphi$ will be associated with relatively slow and fast moving particles respectively. Such a split may not be defined rigidly, but must satisfy certain conditions which we further discuss in section \ref{subsec:III-split}.

Within a set of approximations and assumptions detailed in the subsequent section and appendices, the slow field $\Phi_0$ and $f$ are found to satisfy  a set of self-consistent equations which generalise our study in \cite{Proukakis:2023nmm}:
\begin{subequations}
\begin{eqnarray}
\bullet \quad & &  i \frac{\partial \Phi_0 (x)}{\partial t} = \left( - \frac{1}{2 m a^2} \nabla^2 +
V_{ext}(x) + V_c(x) \right) \Phi_0 (x)  \label{condensateeq1a-mf} \\
& & \hspace{1.5cm} - i R(x) \,\, \Phi_0 (x)  + \xi_1(x)  \label{condensateeq1a-xi1}\\
& & \hspace{1.5cm}  - 2 g \int d^4 x' \Pi^R (x', x) V_{nc} (x') \,\, \Phi_0 (x) + g \Phi_0(x)\,\,\xi_2(x)  \label{condensateeq1a-xi2}
\end{eqnarray}
\end{subequations}
\begin{eqnarray}
\bullet \quad & & \frac{\partial f}{\partial t} + \frac{\mathbf{p}}{m a^2} \cdot\nabla f - \nabla \bigg(V_{ext}(x)
+ V_{nc}(x)\bigg) \cdot \nabla_{\mathbf{p}} f =
\frac{1}{2} (I_a + I_b) \label{particleeq1a}\\
\bullet \quad  & & \frac{1}{4\pi G}\nabla^2 V^{cl} (x) = \frac{m}{a} (n_c (x) + \tilde{n} (x)) - m a^2
\int d^4 x' \Pi^R (x', x) V_{nc} (x') + \frac{1}{2} m a^2 \, \xi_2(x) \label{vcleq1a}
\end{eqnarray}
where the mean field potentials for the coherent and non-coherent parts are respectively
\begin{eqnarray}
V_c (x) &=& m V^{cl} (x) + \frac{g}{a^3} (n_c (x) + 2 \tilde{n} (x)) \;, \label{V-c} \\
V_{nc} (x) &=& m V^{cl} (x) + 2 \frac{g}{a^3} (n_c (x) +
\tilde{n} (x)) \;, \label{V-nc}
\end{eqnarray}
and the number densities for the coherent and non-coherent parts are
\begin{equation}
n_c=|\Phi_0|^2, \quad \tilde{n}=\int \frac{d^3 p}{(2\pi)^3}f(x,\mathbf{p})\,.
\end{equation}
The terms on the right hand side of \eqref{particleeq1a} correspond to collisional terms of this Boltzmann equation and are given by
\begin{eqnarray}
I_a &=& 4 \frac{g^2}{a^6} n_c \int \frac{d^3 p_1 d^3 p_2 d^3 p_3}{(2 \pi)^2} \,
\delta (\varepsilon_c + \varepsilon_{\mathbf{p}_1} -
\varepsilon_{\mathbf{p}_2} - \varepsilon_{\mathbf{p}_3}) \, \delta
(\mathbf{p}_1 - \mathbf{p}_2 - \mathbf{p}_3)\nonumber\\
& & \hspace{1.0cm} \times (\delta (\mathbf{p}_1 - \mathbf{p}) - \delta (\mathbf{p}_2 -
\mathbf{p}) - \delta (\mathbf{p}_3 - \mathbf{p})) ((1 + f_1) f_2 f_3 -
f_1 (1 + f_2) (1 + f_3)) \;, \nonumber\\
I_b &=& 4 \frac{g^2}{a^6} \int \frac{d^3 p_2 d^3 p_3 d^3 p_4}{(2 \pi)^5} \, \delta
(\varepsilon_{\mathbf{p}} + \varepsilon_{\mathbf{p}_2} -
\varepsilon_{\mathbf{p}_3} - \varepsilon_{\mathbf{p}_4}) \, \delta
(\mathbf{p} + \mathbf{p}_2 - \mathbf{p}_3 - \mathbf{p}_4) \nonumber\\
& & \hspace{1.0cm} \times [f_3 f_4 (1 + f) (1 + f_2) - f f_2 (1 + f_3) (1 + f_4)] \label{eqcolls}\,,
\end{eqnarray}
where 
\begin{equation}\label{eq:epsilon_k}
    \varepsilon_\mathbf{p} = \frac{p^2}{2m} + V_{ext} + V_{nc}
\end{equation}
is the single particle energy of the incoherent part and $\varepsilon_c$ corresponds to the energy per particle of the coherent part, given by \cite{ Duine_2001}
\begin{equation}\label{eq:epsilon_c}
    \varepsilon_c = -\frac{1}{2m}\frac{\Phi_0^\ast\nabla^2\Phi_0}{\Phi_0^\ast\Phi_0} + V_{ext} + V_c \;.
\end{equation}
Consistency of this description requires that 
\begin{equation}\label{eq:energy_hierarchy}
\varepsilon_\mathbf{p} \gtrsim \varepsilon_c\,,
\end{equation}
such that the incoherent particles' energies place them above the coherent field in terms of the system's energy spectrum.
The two functions $R$ and $\Pi^R$ appearing within \eqref{condensateeq1a-xi1}-\eqref{condensateeq1a-xi2} are respectively
\begin{eqnarray}
R &=& \frac{g^2}{a^6}
\int \frac{d^3 p_1 d^3 p_2 d^3 p_3}{(2 \pi)^5} \, \delta
(\varepsilon_c + \varepsilon_{\mathbf{p}_1} -
\varepsilon_{\mathbf{p}_2} - \varepsilon_{\mathbf{p}_3}) \, \delta ( \mathbf{p}_1
- \mathbf{p}_2 - \mathbf{p}_3)  \bigg[ f_1 (1 + f_2) (1 + f_3) - (1 + f_1) f_2 f_3 \bigg]\nonumber\\
&=& \frac{1}{4 n_c} \int \frac{d^3 p}{(2 \pi)^3} I_a  \;, \label{rdef2}
\end{eqnarray}

and

\begin{equation}
\Pi^R (x, \mathbf{k}) = \frac{1}{a^3}\int \frac{d^3 p_1 d^3 p_2}{(2 \pi)^3}
\frac{1}{\varepsilon_{\mathbf{k}} + \varepsilon_{\mathbf{p}_2} -
\varepsilon_{\mathbf{p}_1} + i \sigma} \, \delta (\mathbf{k} + \mathbf{p}_2
- \mathbf{p}_1) \bigg[f_1 (1 + f_2) - f_2 (1 + f_1)\bigg] \;, \label{Pi-R}
\end{equation}
where the $i \sigma$ terms arise from integration in the complex plane to obtain the correct causal structure for the retarded object. The quantities $\xi_1$ and $\xi_2$ are Gaussian stochastic forces with correlation functions
\begin{eqnarray}
\langle \xi_1^{\ast} (x) \xi_1 (x') \rangle &=& \frac{i}{2} \Sigma_{(c)}^K (x)\delta(x-x') \;, \label{realcorrelators-a}  \\
\langle \xi_2 (x) \xi_2 (x') \rangle&=&- 2 i \Pi^K (x, x') \;, \label{realcorrelators-b}
\end{eqnarray}
where 
\begin{eqnarray}
\Sigma_{(c)}^K(x) &=& - 2 i \frac{g^2}{a^6}
\int \frac{d^3 p_1 d^3 p_2 d^3 p_3}{(2 \pi)^5} \, \delta
(\varepsilon_c + \varepsilon_{\mathbf{p}_1} -
\varepsilon_{\mathbf{p}_2} - \varepsilon_{\mathbf{p}_3}) \, \delta (\mathbf{p}_1 - \mathbf{p}_2 - \mathbf{p}_3) \nonumber\\
& & \hspace{1.0cm} \times \bigg[f_1 (1 + f_2) (1 + f_3) + (1 +
f_1) f_2 f_3\bigg] \;, \label{sigkdef}
\end{eqnarray}

and
\begin{eqnarray}
\Pi^K (x, \mathbf{k}) &=&  \frac{i}{a^3} \int \frac{d^3 p_1 d^3 p_2}{(2 \pi)^2} \, \delta
(\varepsilon_{\mathbf{k}} + \varepsilon_{\mathbf{p}_2} -
\varepsilon_{\mathbf{p}_1}) \, \delta (\mathbf{k} + \mathbf{p}_2 -
\mathbf{p}_1) \bigg[f_1 (1 + f_2) + f_2 (1 + f_1)\bigg] \;.
\label{pikdef}
\end{eqnarray}

An important point to highlight here is the presence of  two independent noise terms, $\xi_1$ and $\xi_2$, in the dynamical equation for the coherent part \eqref{condensateeq1a-mf}-\eqref{condensateeq1a-xi2}, which are directly related to the two $g^2$-order dynamical terms $-i R$ and $-2 g \int d^4 x' \Pi^R (x', x) V_{nc} (x')$ respectively. 
Looking at the correlations of these noise terms [Eqs.~(\ref{realcorrelators-a})-(\ref{realcorrelators-b})], we highlight that the first one ($\xi_1$) is a complex noise term whose strength is set by the local $\Sigma_{(c)}^K (x)$, while the second term ($\xi_2$) is a real noise term whose strength is set by the non-local $\Pi^K (x, x')$. A quick inspection of their defining equations allows us to observe that $\Sigma^K$ is ``dual" to $R$ while $\Pi^K$ is ``dual" to $\Pi^R$, as expounded upon below.
In both these cases, the evolution of the local value of $\Phi_0$ emerges through the dynamical balancing of the rates of two competing collisional/scattering processes involving a $\Phi_0$ particle.

Let us first consider the $R$ term: 
Its appearance in the form $i \partial \Phi_0 / \partial t = - i R \Phi_0$ implies that such term can directly change the norm of the coherent part, $\Phi_0$.
In fact, looking at (\ref{rdef2}), we see that, at any given time, the local value of $\Phi_0$ is set by the cumulative difference between the local growth terms leading to the creation, and the competing decay terms leading to the loss of coherent $\Phi_0$ particles through energy- and momentum-conserving binary collisions involving incoherent particles, via a collisional process of the schematic form $\varphi + \varphi \leftrightarrow \Phi_0 + \varphi$.
The corresponding noise term amplitude, $\xi_1$, representing fluctuations in such a collisional processes, is thus set by the sum of such incoming and outgoing contributions: this correspondence can be easily seen by a comparison of the integrands appearing within (\ref{rdef2}) and (\ref{sigkdef}). 
We highlight the fact that the presence of the fluctuations $\xi_1$ in (\ref{condensateeq1a-xi1}) implies that the coherent component can grow dynamically from initial fluctuations, even for an initial $\Phi_0 = 0$. As a result, such equations can provide useful insight into the question of the formation of coherent galactic solitonic cores~\cite{Levkov:2018kau,Kirkpatrick:2021wwz,Chen:2021oot}, to be reported elsewhere.
Since this collisional process leads to a change in the number of $\Phi_0$ -- and given that the total particle number (i.e.~the sum of coherent and incoherent particles) is conserved within our formalism  -- this term thus also contributes to a directed collisional integral, $I_a$ in the Boltzmann equation~(\ref{particleeq1a}), through the relation $R=(1/4n_c)\int d^3p/(2 \pi)^3\,I_a$.

Contrary to the above, inspection of \eqref{condensateeq1a-xi2} and the form of \eqref{Pi-R} shows that the process generating the noise term $\xi_2$ only changes the {\em local} value of $\Phi_0$, 
but in a manner which {\em independently} preserves $\int d^3x \, n_c(x)$ and $\int d^3x \, \tilde{n}(x)$.
As such, this $O(g^2)$ term in \eqref{condensateeq1a-xi2} does not in fact lead to an explicit additional collisional integral in the Boltzmann equation~(\ref{particleeq1a}), although {\em locally} $f$ is modified through the $V_{nc}(x)$ term appearing on its left-hand-side.
Such a process does however lead to a change in the energy and momentum of a $\varphi$ particle, associated with the related $f_1(1+f_2)-(1+f_1)f_2 = (f_1-f_2)$ difference factor in (\ref{Pi-R}).
The strength of the noise term associated with such a process is set, through (\ref{pikdef}), by the sum of the two factors, i.e.~$f_1(1+f_2)+(1+f_1)f_2$. 
The process giving rise to such terms can in fact be traced to a `scattering' process involving one coherent and one incoherent particle which exchange energy, but independently preserve their total number, schematically depicted as  $\varphi + \Phi_0 \leftrightarrow \varphi + \Phi_0$.
The $\Pi^R$ contribution appears in the formalism accompanied by the non-condensate mean field potential, $V_{nc}(x)$ [Eq.~(\ref{V-nc})], as evident from (\ref{condensateeq1a-xi2}), (\ref{vcleq1a}). 

Moreover, the $\Pi^R$ and $\Pi^K$ terms also contribute  directly to the Poisson equation, which now also acquires an explicit stochastic noise term associated with such `scattering' processes between coherent and incoherent subsets of the total particle populations comprising the total gravitating mass of the system.

Interestingly -- and in addition to all the non-stochastic models arising as limiting cases of our theory as previously discussed in ~\cite{Proukakis:2023nmm} -- the above set of stochastic equations can be viewed as a formal methodological extension of related stochastic models in different physical systems. We briefly mention two relevant cases here:
Firstly, in the context of ultracold atomic gases~\cite{Proukakis13Quantum} (where there is no time dependent scale factor, i.e.~$a\rightarrow1$), and based on distinct formalisms, subsets of \eqref{condensateeq1a-mf}, \eqref{condensateeq1a-xi1}, \eqref{condensateeq1a-xi2} and \eqref{particleeq1a} have been previously derived: Specifically, a coupled stochastic Gross-Pitaevskii-Boltzmann system of equations \eqref{condensateeq1a-mf}, \eqref{condensateeq1a-xi1}, \eqref{particleeq1a} was formulated in~\cite{stoof1999coherent,Duine_2001,stoof_dynamics_2001}, but without the inclusion of the `scattering' terms of \eqref{condensateeq1a-xi2} related to $\Pi^R$ and $\xi_2$. 
Such scattering terms were in fact explicitly included 
in~\cite{Gardiner_2003,bradley2006stochasticgrosspitaevskiiequationiii,PhysRevA.77.033616,Blakie-AdvPhys-2008} (from where the `scattering' terminology has been borrowed), but such equation appeared under the assumption of a thermal equilibrium (and the further approximation of the Bose-Einstein distribution by a Rayleigh-Jeans distribution for highly-occupied modes), thus dropping the Boltzmann equation \eqref{particleeq1a}, while simultaneously treating \eqref{condensateeq1a-xi1} in an approximate manner.
Such a relation is discussed in more detail in our accompanying work focussing directly on ultracold gases~\cite{Proukakis-coldatom-stochastic}. 
Moreover, equations \eqref{particleeq1a}, \eqref{vcleq1a} have been discussed in the context of collisionless plasma physics, but only in the presence of incoherent particles, i.e.~setting $\Phi_0=n_c=0$ and $I_a = I_b =0$ in them~\cite{kamenev_2011}. 


\section{Step-by-step Recipe for obtaining the dynamical equations} \label{Steps}

In this section we describe in more detail how equations \eqref{condensateeq1a-mf}, \eqref{condensateeq1a-xi1}, \eqref{condensateeq1a-xi2}, \eqref{particleeq1a}, \eqref{vcleq1a} can be derived, starting from the action \eqref{eqini}. To aid readability, we focus on mapping out the basic steps and relegate some technical details to the appendices. For convenience, we will work in flat space to make the construction simpler, and at the end we will make use of appropriate transformations to express our results for an expanding universe.

\subsection{Step I: Schwinger-Keldysh action}

We will work in the Schwinger-Keldysh formalism \cite{1986PhRvD..33..444J, 1988PhRvD..37.2878C, Stoof-PhysRevLett.78.768, stoof1999coherent,RammerBook,kamenev_2011}, where the evolution of a non-equilibrium density matrix involves a doubling of the Hilbert space, with one half of the degrees of freedom evolving forward in time and the other half evolving backwards. In practice, we double the field degrees of freedom in our starting action, denoting them by $\psi^+$ and $\psi^-$, depending on whether they are on the forward or backward contour respectively, and further satisfying  $\psi^+(t_f)=\psi^-(t_f)$ at some final time $t_f$ in the far future. The same is done for the field $V$, and we write the corresponding Schwinger-Keldysh action as $S = S[\psi^+,V^+] - S[\psi^-,V^-]$, where the minus sign comes from the fact that we are going backwards in the temporal integration. With this, our Schwinger-Keldysh action explicitly reads
\begin{eqnarray}
S &=& \int d t d^3 \mathbf{r} \bigg( i \psi^{\ast +} \dot{\psi}^+ +
\frac{1}{2 m} \psi^{\ast +} \nabla^2 \psi^+ - \frac{g}{2} (\psi^{+ \ast}\psi^+)^2 -
V_{ext} \psi^{+ \ast}\psi^+ + \frac{1}{8\pi G} V^+ \nabla^2 V^+ - m V^+ \psi^{+ \ast}\psi^+ \nonumber\\
& & - i \psi^{\ast -} \dot{\psi}^- -
\frac{1}{2 m} \psi^{\ast -} \nabla^2 \psi^- + \frac{g}{2} (\psi^{- \ast}\psi^-)^2 +
V_{ext} \psi^{- \ast}\psi^- - \frac{1}{8\pi G} V^- \nabla^2 V^- + m V^- \psi^{- \ast}\psi^- \bigg) \;.
\end{eqnarray}
We then use a Keldysh rotation, defining
\begin{eqnarray}
    \psi^{cl} = \frac{\psi^+ + \psi^-}{\sqrt{2}}\,, \quad\psi^{q} = \frac{\psi^+ - \psi^-}{\sqrt{2}} \\
    V^{cl} = \frac{V^+ + V^-}{2}\,, \quad V^{q} = \frac{V^+ - V^-}{2}
\end{eqnarray}
where the indices $cl$ and $q$ stands for the so called ``classical" and ``quantum" components in this formalism, the latter signifying fluctuations around the on-shell solution. With these transformations the action reads
\begin{eqnarray}
\label{actionkeldysh}
S &=& \int d t d^3 \mathbf{r} \bigg( \psi^{q \ast} \left( i \partial_t +
\frac{1}{2 m} \nabla^2 - V_{ext} \right) \psi^{cl} +
\psi^{cl \ast} \left( i \partial_t + \frac{1}{2 m} \nabla^2
- V_{ext} \right) \psi^q \nonumber\\
& & - \frac{g}{2} \bigg( \psi^{q \ast} (|
\psi^{cl} |^2 + | \psi^q |^2) \psi^{cl} + \psi^{cl \ast}
(| \psi^{cl} |^2 + | \psi^q |^2) \psi^q \bigg) + \frac{1}{4\pi G}\left( V^q
\nabla^2 V^{cl} + V^{cl} \nabla^2 V^q \right) \nonumber\\
& & - m
V^{cl} (\psi^{q \ast} \psi^{cl} + \psi^{cl \ast} \psi^q)
- m V^q (\psi^{cl \ast} \psi^{cl} + \psi^{q \ast} \psi^q)
\bigg) \;.
\end{eqnarray}

\subsection{Step II: Split the bosonic field into slow and fast components}\label{subsec:III-split}

Following the basic premise of our work, i.e.~that the we can separate fast-moving from slow-moving modes, we split the classical and quantum fields into a slow part, taken to represent modes belonging to the
coherent part, and a fast one, representing the non-coherent part or the ``particles", as
\begin{equation}
\label{splitting}
\psi^{cl} = \Phi_0 +
\varphi, \qquad \psi^q = \Phi^q + \varphi^q \;.
\end{equation}
The precise energy at which this split is made, designating some modes to the coherent part and the rest to the incoherent part, is left unspecified for the moment and our formal expressions do not depend on the precise choice.

\subsubsection{Qualitative Discussion of ``Slow/Coherent'' vs.~``Fast/Incoherent'' Component Split \label{sec:cut}}

The energy, $\varepsilon^{cut}$, at which this split is made should be consistent with the following broad requirements:
\begin{itemize}
\item It cannot be lower than the energy $\varepsilon_c^{po}$ of the field configuration's Penrose-Onsager mode~\cite{2022arXiv221102565L}, corresponding to the lowest-energy mode of the system and describing the purely condensed part of the system. The Penrose-Onsager mode is characterized by the absence (or at least sub-dominance) of phase fluctuations and has
\begin{equation}
    \varepsilon_c^{po} = -\frac{1}{2m}\frac{\nabla^2\sqrt{n_{c}^{po}}}{\sqrt{n_{c}^{po}}} + V_{ext} + V_c \;.
\end{equation}
Setting $\varepsilon^{cut}$ lower than this would eliminate the ability to describe the Penrose-Onsager mode and therefore condensation; hence $\varepsilon^{cut}\gtrsim \varepsilon_c^{po}$.
\item It cannot be higher than the energy, $\varepsilon_{n_k \sim 1}$, where $|\psi(k)|^2 \sim 1$. This ensures that the occupation numbers of all modes making up $\Phi_0$ are high enough to justify a continuum field description for it.\footnote{Note that for fuzzy dark matter boson masses $m \gtrsim 10 \, eV$ the occupation numbers in dark matter halos can indeed become less than ${\cal O}(1)$ particle per mode \cite{2021A&ARv..29....7F} and a description in terms of a phase space density $f$ is more than merely a convenient choice, becoming in fact a necessity.} 
\end{itemize}
In other words, a minimum condition for the location of the cutoff must be\footnote{See also related discussion of such issues in the cold atom community~\cite{Blakie-AdvPhys-2008}.}
\begin{equation}
\varepsilon_c^{po} \lesssim \varepsilon^{cut} \lesssim 
\varepsilon_{n_k \sim 1}
\label{conditions}
\end{equation}

In principle, the cutoff could be placed anywhere between these two bounds provided that a fast-slow separation can be maintained. This separation can be quantified by demanding that the field $\Phi_0(x)$ shows a slower spatio-temporal variation than the distance over which $\langle\varphi(x)\varphi^{\ast}(x')\rangle = G^K(x,x')$ is non-negligible. 
Writing  
\begin{equation}
    G^K(x,x') = G^K\left(\frac{x+x'}{2},x-x'\right) = \underset{p}{\sum} e^{i p (x - x')} G \left( \frac{x + x'}{2}, p \right) \,,
\end{equation}
where we have also defined the Wigner transform in the second equation, our fast-slow separation is valid if there exists a maximum spatiotemporal variation $\Delta x_{max} = \left(x-x'\right)_{max}\,$, such that the rapidly-varying fluctuations encoded in $G^K\left(\frac{x+x'}{2},x-x'\right)$ decay to zero, i.e.~$G^K\left(\frac{x+x'}{2},\Delta x_{max}\right) \simeq 0 $, while the coherent field $\Phi_0$ does not appreciably change, i.e.~$\Phi_0\left(\frac{x+x'}{2}+\Delta x_{max}\right) \simeq  \Phi_0\left(\frac{x+x'}{2}\right)$.
This implies that
\begin{equation}\label{eq:gradient_bound}
    \frac{\left|\nabla\Phi_0\right|}{\Phi_0}\lesssim\frac{1}{\left|\Delta 
 \mathbf{x}_{max}\right|} = \frac{\left|\mathbf{p}_{min}\right|}{2\pi}\,,\quad \frac{\partial_t\Phi_0}{\Phi_0}\lesssim\frac{1}{\Delta t_{max}} = \varepsilon_{\mathbf{p}}^{min} \;,
\end{equation}
where $\left(\varepsilon_{\mathbf{p}}^{min},\,\mathbf{p}_{min}\right)$ is the smallest 4-momentum for which the Wigner transform $G \left( \frac{x + x'}{2}, p \right)$ is appreciable, and corresponds to a low bound for the momenta of the particles making up the incoherent part. 
Note that the latter expression appearing in (\ref{eq:gradient_bound}) can be understood as a hierarchy in terms of the energy densities per particle, $\partial_t \Phi_0 / \Phi_0 \sim  \varepsilon_c \lesssim \varepsilon_\mathbf{p}$, which is consistent with \eqref{eq:energy_hierarchy}.  This can be re-expressed in the form
\begin{equation}\label{eq:momentum_bound}
    |\mathbf{p}|>\sqrt{-\frac{\Phi_0^\ast\nabla^2\Phi_0}{n_c} - 2gm n_c} \,\,\,,
\end{equation}
for the incoherent particle 3-momenta, translating into a lower bound for the momenta in all the phase space integrals of section \ref{Equations}. 
Note that this limit is generally higher than the minimal condition of the coherent part comprised only of the Penrose-Onsager mode.
We argue that the coherent field $\Phi_0$ can generally encompass both the Penrose-Onsager (condensate) mode and a population of adjacent ``quasi-coherent'' modes, which exhibit non-negligible spatio-temporal coherence.\footnote{To ensure we fully incorporate the condensate and dominant adjacent wave-like modes in future numerical simulations, we would propose trialing a cutoff a few times the Penrose-Onsager mode and investigating the effect of varying the cutoff.
Due to the self-consistent dynamical description of both ``coherent'' and ``incoherent'' modes, we would expect the results of a full coupled simulation of (\ref{condensateeq1a-mf}) - (\ref{vcleq1a}) to be largely insensitive to the exact location of the cutoff, provided both conditions (\ref{conditions}) are fully satisfied. We intend to test this in a future work.} As such, ``incoherent particles" are associated with momenta higher than the typical inverse fluctuation scale of the coherent field.
Note that relation \eqref{eq:gradient_bound} is assumed in the manipulations presented in the Appendices (see for example \eqref{eq:local_approx}) leading to the equations of section \ref{Equations}.
%


\subsubsection{Actions for the slow and fast modes}

We will further work in an approximation where we consider the gravitational potential $V$ and
the external potential $V_{ext}$ as slow varying quantities only.
The most important condition that we will highlight for our slow fields $\Phi$ and $V$ is that we will keep terms only up to second order in their quantum components $\Phi^q$ and $V^q$: the first order terms will provide a semiclassical description of the coherent part $\Phi_0$ and the potential $V$, while the next order will describe stochastic fluctuations. In contrast, we will keep all higher orders in $\varphi^q$, the quantum component of the field's fast part, tracking all its fluctuations. This will be necessary as we will be integrating out the the fast fields to obtain effective equations for the slow fields.
%
%

With such considerations, the action \eqref{actionkeldysh} now reads
\begin{equation}
S = S_{0 (fast)} + S_{0 (slow)} + S_I \;,
\end{equation}
where
\begin{eqnarray}
S_{0 (fast)} &=& \int d^4 x \left(\begin{array}{cc}
  \varphi^{\ast} \: , & \varphi^{q \ast}
\end{array}\right) \left(\begin{array}{cc}
  0 & i \partial_t + \frac{1}{2 m} \nabla^2 - V_{ext}\\
  i \partial_t + \frac{1}{2 m} \nabla^2 - V_{ext} & 0
\end{array}\right) \left(\begin{array}{c}
  \varphi\\
  \varphi^q
\end{array}\right) \;,\\
S_{0 (slow)} &=& \int d^4 x \left(\begin{array}{cc}
  \Phi_0^{\ast} \:, & \Phi^{q \ast}
\end{array}\right) \left(\begin{array}{cc}
  - m V^q & i \partial_t +\frac{1}{2 m} \nabla^2-V_{ext}- V_{c}\\
  i \partial_t + \frac{1}{2 m} \nabla^2 -V_{ext} - V_{c} & 0
\end{array}\right) \left(\begin{array}{c}
  \Phi_0\\
  \Phi^q
\end{array}\right)\nonumber\\
& & + \frac{1}{4\pi G}\int d^4 x \left(\begin{array}{cc}
  V^{cl} \:,& V^q
\end{array}\right) \left(\begin{array}{cc}
  0 & \nabla^2\\
  \nabla^2 & 0
\end{array}\right) \left(\begin{array}{c}
  V^{cl}\\
  V^q
\end{array}\right) \nonumber\\
& & - m \int d^4 x V^q \langle \varphi^{\ast} \varphi \rangle \;, 
\end{eqnarray}
where we have defined a mean field potential $V_{c}$ for the slow component $\Phi$  as
\begin{equation}
\label{Vc}
V_c(x) = m V^{cl}(x) + g \bigg( \frac{1}{2} |
\Phi_0(x) |^2 + \langle \varphi^{\ast}(x) \varphi(x) \rangle \bigg) \;.
\end{equation}
The interaction action $S_I$ is defined by two pieces:
\begin{equation}
\label{Si}
S_I = S_{g} + S_{m} \;, 
\end{equation}
where
\begin{equation}
S_{g} = S_{g}^{(2)} + S_{g}^{(3)} + S_{g}^{(4)} \;, 
\end{equation}
with
\begin{eqnarray}
S_{g}^{(2)} &=& - \frac{g}{2} \int d^4 x \bigg( 2 \Phi_0
\Phi_0^{\ast} \varphi \varphi^{q \ast} + 2 \Phi_0 \Phi^{q \ast} \varphi
\varphi^{\ast} + 2 \Phi^{q \ast} \Phi_0 \varphi^{q \ast} \varphi^q + 2 \Phi^{q
\ast} \Phi^q \varphi^{q \ast} \varphi + 2 \Phi_0^{\ast} \Phi^q \varphi^{\ast}
\varphi + 2 \Phi_0^{\ast} \Phi_0 \varphi^{\ast} \varphi^q \nonumber\\
& & + 2 \Phi^q \Phi^{q
\ast} \varphi^q \varphi^{\ast} + 2 \Phi^q \Phi_0^{\ast} \varphi^q \varphi^{q
\ast} - 2 \Phi^{q \ast} \langle \varphi^{\ast} \varphi \rangle \Phi_0 - 2
\Phi_0^{\ast} \langle \varphi^{\ast} \varphi \rangle \Phi^q \bigg) \;, \label{Si2}\\
S_{g}^{(3)} &=& - \frac{g}{2} \int d^4 x \bigg( 2 \Phi_0 \varphi
\varphi^{q \ast} \varphi^{\ast} + \Phi_0^{\ast} \varphi^{q \ast} \varphi
\varphi + \Phi^{q \ast} \varphi^{\ast} \varphi \varphi + 2 \Phi^{q \ast}
\varphi^{q \ast} \varphi^q \varphi + \Phi_0 \varphi^{q \ast} \varphi^{q \ast}
\varphi^q + \Phi^q \varphi^{q \ast} \varphi^{q \ast} \varphi \nonumber\\
& & + 2 \Phi_0^{\ast}
\varphi^{\ast} \varphi \varphi^q + \Phi^q \varphi^{\ast} \varphi^{\ast}
\varphi + \Phi_0 \varphi^{\ast} \varphi^{\ast} \varphi^q + 2 \Phi^q \varphi^q
\varphi^{\ast} \varphi^{q \ast} + \Phi^{q \ast} \varphi^{\ast} \varphi^q
\varphi^q + \Phi_0^{\ast} \varphi^{q \ast} \varphi^q \varphi^q \bigg) \;, \label{Si3}\\
S_{g}^{(4)} &=& - \frac{g}{2} \int d^4 x \bigg(\varphi^{q \ast}
\varphi^{\ast} \varphi \varphi + \varphi^{q \ast} \varphi^{q \ast} \varphi^q
\varphi + \varphi^{\ast} \varphi^{\ast} \varphi \varphi^q + \varphi^{\ast}
\varphi^{q \ast} \varphi^q \varphi^q \bigg) \;,  \label{Si4}
\end{eqnarray}
and where
\begin{eqnarray}
S_{m} = - m \int d^4 x \ V^{cl} (\varphi^{q \ast}
\varphi + \varphi^{\ast} \varphi^q) - m \int d^4 x \ V^q
(\varphi^{\ast} \varphi - \langle \varphi^{\ast} \varphi \rangle + \varphi^{q
\ast} \varphi^q) \;. \label{Sia2}
\end{eqnarray}

Important comments are in order. First, after the splitting in slow and fast fields using \eqref{splitting}, we have discarded all terms that violate energy-momentum conservation. Thus, terms with a single $\varphi$ field do not contribute. This class of objects are: terms with one
slow field $\Phi$ and one fast $\varphi$, terms with three slow $\Phi$ and one fast $\varphi$, and
terms with one slow $\Phi$, one slow $V$ and one fast $\varphi$.
Second, we have placed the slow terms $V^{cl}$ and $| \Phi_0 |^2$ inside the coherent field hamiltonian as parts of the mean field potential $V_{c}$ since they are composed just by classical fields. This means that combinations involving both slow and fast fields are kept in $S_I$. This consideration departs from the procedures followed in \cite{kamenev_2011} and in our previous work \cite{Proukakis:2023nmm}. Considering only up to order one in the quantum fields leads to no difference in doing this. However, this step becomes important in order to get correct physical results at higher order in the quantum fields. Also, for later convenience that will become apparent in the next section, we have added the term $g \langle \varphi^{\ast} \varphi \rangle$ in the mean field potential and subtracted it in the interacting part $S_{g}^{(2)}$. In the same way, we have placed a term $m  V^q\langle \varphi^{\ast} \varphi \rangle$ in the slow action and subtracted it in $S_{g}^{(4)}$. These introduced bracketed objects are two point functions, corresponding to the dressed propagator for the fast $\varphi$-particles. At this point, we observe that the hamiltonian for the incoherent (fast) $\varphi$-particles is just the free one plus the external potential $V_{ext}$. However, we will see later that it will be modified when we compute the corresponding self-energies.

\subsection{Step III: Effective action for the slow fields via perturbation theory in the generating functional}

Now, we turn our attention to the generating functional
\begin{equation}
Z = \int \mathcal{D} [\Phi V \varphi] e^{i S_{0 (slow)} + i S_{0
(fast)} + i S_I} \;, 
\end{equation}
which we expand in powers of the interacting part as
\begin{eqnarray}
Z \approx \int \mathcal{D} [\Phi V] e^{i S_{0 (slow)}} \left( 1 + \frac{\int
\mathcal{D} [\varphi] e^{i S_{0 (fast)}} (i S_{g} + i
S_{m})}{\int \mathcal{D} [\varphi] e^{i S_{0 (fast)}}} +
\frac{1}{2} \frac{\int \mathcal{D} [\varphi] e^{i S_{0 (fast)}} (i
S_{g} + i S_{m})^2}{\int \mathcal{D} [\varphi] e^{i S_{0
(fast)}}} \right) \int \mathcal{D} [\varphi] e^{i S_{0 (fast)}} \;. 
\end{eqnarray}
In the Schwinger-Keldysh formalism we have that 
\begin{equation}
\int \mathcal{D} [\varphi] e^{i S_{0 (fast)}} = 1 \;, 
\end{equation}
and therefore,
\begin{equation}
Z \approx \int \mathcal{D} [\Phi V] e^{i S_{0 (slow)}} \left( 1 + i \langle
S_{g} \rangle + i \langle S_{m} \rangle - \frac{1}{2} \langle
S_{g}^2 \rangle - \frac{1}{2} \langle S_{m}^2 \rangle - \langle
S_{g} S_{m} \rangle \right) \;. 
\end{equation}
Due to our addition and subtraction of terms in the mean field potential and interaction action $S_I$, a straightforward computation shows that $\langle S_{g} \rangle = \langle S_{m} \rangle = 0$ and as a consequence we can rewrite the expansion as an exponential again, such that we have
\begin{equation}
Z \approx \int \mathcal{D} [\Phi V] e^{i S_{eff}} \;, 
\end{equation}
where we have defined an effective action $S_{eff}$ as
\begin{equation}
\label{effaction0}
S_{eff} = S_{0 (slow)} + \frac{1}{2} i \langle S_{g}^2
\rangle + \frac{1}{2} i \langle S_{m}^2 \rangle + i \langle
S_{g} S_{m} \rangle \;. 
\end{equation}

When computing the bracketed quantities,  the below Schwinger-Keldysh \emph{bare propagators} for the fast field make a repeated appearance 
\begin{eqnarray}
\langle \varphi(x) \varphi^{\ast}(x') \rangle &=& i G^K_{0(nc)} (x, x') \;, \\
\langle \varphi(x) \varphi^{q \ast}(x')  \rangle &=& i G^R_{0(nc)} (x, x') \;, \\
\langle \varphi^q(x)  \varphi^{\ast}(x') \rangle &=& i G^A_{0(nc)} (x, x') \;, 
\end{eqnarray}
where $G_{0(nc)}^{R(A)}$ stands for the retarded (advanced) propagator and $G_{0(nc)}^K$ corresponds to the Keldysh propagator of the fast fields. On the other hand, we can relate these bare propagators to the \emph{dressed propagator} 
\begin{equation}
\label{proppart}
G_{(nc)}=\left(\begin{array}{cc}
  G_{(nc)}^K & G_{(nc)}^R\\
  G_{(nc)}^A & 0
\end{array}\right)
\end{equation}
through the Schwinger-Dyson equation for the two point correlation function
\begin{equation}\label{SDeq1}
G_{(nc)} = G_{0(nc)}+G_{0(nc)} \otimes\Sigma_{(nc)}\otimes G_{(nc)} \;, 
\end{equation}
where the product sign $\otimes$ implies appropriate convolutions in spacetime and also multiplication in the Keldysh matrix space. The object $\Sigma_{(nc)}(x,x')$, also a matrix with the Keldysh structure 
\begin{equation}
\Sigma_{(nc)}=\left(\begin{array}{cc}
   0 & \Sigma_{(nc)}^A\\
  \Sigma_{(nc)}^R & \Sigma_{(nc)}^K
\end{array}\right)\,,
\end{equation}
corresponds to the self-energy of the particles and is of order $g$ and higher. This means that for the objects in equation \eqref{effaction0} we can approximate
\begin{equation}
G_{0(nc)}\approx G_{(nc)} 
\end{equation}
in our expressions. In consequence, any propagator appearing in the action will be taken to correspond to the dressed one.

We will be using the Popov approximation \cite{popov1987functional,PhysRevB.53.9341,Hutchinson:1997zz} where
\begin{equation}
\langle \varphi\varphi\rangle \simeq \langle \varphi^\ast\varphi^\ast\rangle \simeq \langle \varphi^q\varphi^q\rangle \simeq\langle \varphi^{q\ast}\varphi^{q\ast}\rangle \simeq 0 \;, 
\end{equation}
and the following identities for the Schwinger-Keldysh propagators \cite{kamenev_2011} 
\begin{eqnarray}
G^R (x,
x) + G^A (x, x) = 0 \;, \\
G^{R (A)} (x, x') G^{R (A)} (x', x) = 0 \;,  \\ 
G^R (x, x')
G^A (x, x') = 0 \;. 
\end{eqnarray} 
With the above and after a long computation the effective action reads
\begin{eqnarray}
\label{effaction0b}
S_{eff} &=& \int d^4 x \left(\begin{array}{cc}
  \Phi_0^{\ast} (x) \: , & \Phi^{q \ast} (x)
\end{array}\right) \left(\begin{array}{cc}
  0 & i \partial_t - H_c\\
  i \partial_t - H_c & 0
\end{array}\right) \left(\begin{array}{c}
  \Phi_0 (x)\\
  \Phi^q (x)
\end{array}\right) \nonumber\\
& & + \int d^4 x' d^4 x \left(\begin{array}{cc}
  \Phi_0^{\ast} (x') \: , & \Phi^{q \ast} (x')
\end{array}\right) \left(\begin{array}{cc}
  0 & - \Sigma_{(c)}^A (x', x)\\
  - \Sigma_{(c)}^R (x, x') & 0
\end{array}\right) \left(\begin{array}{c}
  \Phi_0 (x)\\
  \Phi^q (x)
\end{array}\right)\nonumber\\
& & + \frac{1}{2\pi G} \int d^4 x \, V^q (x) \nabla^2 V^{cl} (x) - m \int d^4 x \, V^q
(x) \bigg(| \Phi_0 (x) |^2 + \langle \varphi^{\ast} (x) \varphi (x) \rangle \bigg)\nonumber\\
& & + 2 m \int d^4 x' d^4 x \, V^q (x) \Pi^R (x', x) V_{nc}(x') \nonumber\\
& & + \int d^4 x' d^4 x \, \bigg(g (\Phi^{q \ast} \Phi_0 + \Phi_0^{\ast} \Phi^q)
(x') + m V^q (x') \bigg) \Pi^K (x', x) \bigg(g (\Phi^{q \ast} \Phi_0 +
\Phi_0^{\ast} \Phi^q) (x) + m V^q (x) \bigg) \nonumber\\
& & - \int d^4 x' d^4 x \, \Phi^{q \ast} (x') \Sigma_{(c)}^K (x', x) \Phi^q (x) \;, 
\end{eqnarray}
where we have denoted by
\begin{equation}
H_c = -\frac{1}{2 m} \nabla^2 + V_{ext}+ V_c(x) - 2 g \int d^4 x' \Pi^R (x', x) V_{nc}(x') \;, 
\end{equation}
with $V_c$ defined as in \eqref{Vc} and
\begin{equation}
\label{Vnc}
V_{nc}(x)=m V^{cl} (x) +
g \bigg(| \Phi_0 (x) |^2 + \langle \varphi^{\ast} (x) \varphi (x)
\rangle \bigg)\,.
\end{equation}
The sub-index $nc$ indicates here, as we will see soon, that it corresponds to the mean-field potential for the non-coherent, fast particles. In the above expressions we have defined
\begin{eqnarray}
\Pi^K (x', x) &=& - \frac{i}{2} \bigg(G_{(nc)}^K (x', x) G_{(nc)}^K (x, x') + G_{(nc)}^A (x', x) G^R (x,
x') + G_{(nc)}^R (x', x) G_{(nc)}^A (x, x') \bigg) \;, \label{PiK0}\\
\Pi^R (x', x) &=& - \frac{i}{2} \bigg(G_{(nc)}^R (x', x) G_{(nc)}^K (x, x') + G_{(nc)}^K (x', x) G_{(nc)}^A (x,
x') \bigg) \;, \label{PiR0}\\
\Sigma_{(c)}^K (x', x) &=& - \frac{g^2}{2} \bigg( G_{(nc)}^K (x', x) G_{(nc)}^K (x,
x') G_{(nc)}^K (x, x') + G_{(nc)}^K (x', x) G_{(nc)}^R (x, x') G_{(nc)}^R (x, x') \nonumber\\
& & + G_{(nc)}^K (x', x) G_{(nc)}^A (x,
x') G_{(nc)}^A (x, x') + 2 G_{(nc)}^K (x, x') G_{(nc)}^A (x', x) G_{(nc)}^R (x, x') \nonumber\\
& & + 2 G_{(nc)}^K (x, x') G_{(nc)}^R
(x', x) G_{(nc)}^A (x, x') \bigg) \;, \label{sigmaK0}\\
\Sigma_{(c)}^R (x', x) &=& - \frac{g^2}{2} \bigg( G_{(nc)}^R (x', x) G_{(nc)}^R (x',
x) G_{(nc)}^A (x, x') + G_{(nc)}^K (x', x) G_{(nc)}^K \left( x', x \right) G_{(nc)}^A (x, x') \nonumber\\
& & + 2 G_{(nc)}^R (x',
x) G_{(nc)}^K (x', x) G_{(nc)}^K (x, x') \bigg) \;, \label{sigmaR0}\\
\Sigma_{(c)}^A (x', x) &=& - \frac{g^2}{2} \bigg( G_{(nc)}^R (x', x) G_{(nc)}^K (x,
x') G_{(nc)}^K (x, x') + G_{(nc)}^A (x, x') G_{(nc)}^A \left( x, x' \right) G_{(nc)}^R (x', x) \nonumber\\
& & + 2 G_{(nc)}^K
(x', x) G_{(nc)}^A (x, x') G_{(nc)}^K (x, x') \bigg) \;. \label{sigmaA0}
\end{eqnarray}

To simplify the notation a little bit, we will use the fact that for any Keldysh propagator $i G^K (x, x) = 2 n_B + 1$, where $n_B$
is the bosonic occupation number. Note that $| \Phi_0 |^2$ can also be seen as the Keldysh propagator for the coherent part. This means that
\begin{equation}
\int d \mathbf{r} | \Phi_0 |^2 = 2 \int d \mathbf{r} n_c = 2 N_c \;, 
\end{equation}
where $n_c$ corresponds to the number density of the coherent (slow) part which, when integrated in space, gives the total number of particles $N_c$ assigned to the coherent part. Here we have neglected the factor 1 in the Keldysh expression since this term gives a vacuum energy which can be subtracted by means of a renormalisation of the field. We also observe that the normalization of the $\Phi_0$ field corresponds in the Schwinger-Keldysh formalism to twice the number of constituent particles, To normalise to a particle number $N_c$, as is usual, we will redefine $\Phi_0 \rightarrow \sqrt{2} \Phi_0$ at the end of the computations. Meanwhile, we will replace in our expressions
\begin{equation}
\label{condensatedens}
n_c = \frac{1}{2} | \Phi_0 |^2\,.
\end{equation}
For the non-coherent (fast) part, we use the usual way of writing any Keldysh function as \cite{kamenev_2011,Altland,Sieberer:2015svu}
\begin{equation}
\label{gkeldF}
G^K = G^R \circ F - F \circ G^A \;, 
\end{equation}
where the symbol $\circ$ represents convolution and $F (x, x')$ is a Hermitian matrix in the spacetime domain. Its Wigner
transform (see \eqref{wignertra}) is $\tilde{F} (x, \mathbf{p})$\footnote{Note that $x$ in $\tilde{F}$ corresponds to the central point coordinates i.e.~$({x+x'})/{2}\rightarrow x$ and $\mathbf{p}$ is the momentum related with the Fourier transform of the relative coordinate $x-x'$.} and here any energy dependence in
$\tilde{F}$ is disregarded assuming that it is a slower function than $G^R -
G^A$ \cite{kamenev_2011}. This term is related to the Keldysh propagator as
\begin{equation}
\tilde{F} (x, \mathbf{p}) = i G^K (x, \mathbf{p}) \;, 
\end{equation}
and with it we can set the following relation
\begin{equation}
\label{defspartprop}
\langle \varphi^{\ast} (x) \varphi (x) \rangle = i G_{(nc)}^K (x, x) =
\underset{\mathbf{p}}{\sum} \tilde{F} (x, \mathbf{p}) =
\underset{\mathbf{p}}{\sum} (2 f (x, \mathbf{p}) + 1) \;, 
\end{equation}
where $f$ is the distribution function associated with the occupation number of the mode with momentum $\mathbf{p}$. Hence, we can define the non-coherent number density
\begin{equation}
\label{quasipartdens}
\tilde{n}=\underset{\mathbf{p}}{\sum} f \;, 
\end{equation}
and neglecting the last sum over $1$ in \eqref{defspartprop}, since as in the coherent case it gives a vacuum energy which can be
subtracted by means of a renormalisation, we can write
\begin{equation}
\label{ntildeexpect}
2 \tilde{n} = \langle \varphi^{\ast} (x) \varphi (x) \rangle \;. 
\end{equation}
In addition to this, to get rid of the non-locality of the terms containing the coherent self-energies in equation \eqref{effaction0b}, we can take a Wigner transform; for details on that computation and the terms arising, we refer to appendix \ref{appWig}.

The interpretation of the effective action and the resulting equations of motion is further facilitated if the action is transformed to an expression linear in the quantum fields $\Phi^q$ and $V^q$. This is achieved by a Hubbard-Stratonovich transformation of the last two lines. With these considerations, the effective action becomes 
\begin{eqnarray}
\label{sefftotal}
S_{eff} &=& \int d^4 x \left(\begin{array}{cc}
  \Phi_0^{\ast} & \Phi^{q \ast}
\end{array}\right) \left(\begin{array}{cc}
  0 & i \partial_t - H_c - i R\\
  i \partial_t - H_c + i R & 0
\end{array}\right) \left(\begin{array}{c}
  \Phi_0\\
  \Phi^q
\end{array}\right) \nonumber\\
& & + \frac{1}{2\pi G} \int d^4 x \,\,V^q (x) \nabla^2 V^{cl} (x) - 2 m \int d^4 x\,\,
V^q (x) (n_c (x) + \tilde{n} (x)) \nonumber\\
& & + 2 m \int d^4 x' d^4 x \,\, V^q (x) \Pi^R (x', x) V_{nc}(x') \nonumber\\
& & + \int d^4 x' d^4 x \,\, \xi_1^{\ast} (x') (\Sigma_{(c)}^K (x,x'))^{- 1} \xi_1 (x) \nonumber\\
& & - \int d^4 x \,\, (\xi_1^{\ast} \Phi^q + \Phi^{q \ast} \xi_1) \nonumber\\
& & - \frac{1}{4} \int d^4 x' d^4 x \, \,\xi_2 (x') (\Pi^K (x', x))^{- 1} \xi_2 (x) \nonumber\\
& & - g \int d^4 x \,\, \xi_2 (\Phi^{q \ast} \Phi_0 + \Phi_0^{\ast} \Phi^q) -
m \int d^4 x \,\,\xi_2 V^q \;, 
\end{eqnarray}
where we recall that
\begin{equation}
H_c = - \frac{1}{2 m} \nabla^2 + V_{ext}(x) + V_c(x) - 2 g \int d^4 x' \,\,\Pi^R (x', x) V_{nc}(x') \;, 
\end{equation}
and
\begin{eqnarray}
R &=& g^2
\int \frac{d^3 p_1 d^3 p_2 d^3 p_3}{(2 \pi)^5} \delta
(\varepsilon_c + \varepsilon_{\mathbf{p}_1} -
\varepsilon_{\mathbf{p}_2} - \varepsilon_{\mathbf{p}_3}) \delta (\mathbf{p}_1
- \mathbf{p}_2 - \mathbf{p}_3) \nonumber\\
& & \times \bigg[ f_1 (1 + f_2) (1 + f_3) - (1 + f_1) f_2 f_3 \bigg] \;, \label{rdef}\\
\Sigma_{(c)}^K(x) &=& - 2 i g^2
\int \frac{d^3 p_1 d^3 p_2 d^3 p_3}{(2 \pi)^5} \delta
(\varepsilon_c + \varepsilon_{\mathbf{p}_1} -
\varepsilon_{\mathbf{p}_2} - \varepsilon_{\mathbf{p}_3}) \delta (\mathbf{p}_1 - \mathbf{p}_2 - \mathbf{p}_3) \nonumber\\
& & \bigg[f_1 (1 + f_2) (1 + f_3) + (1 +
f_1) f_2 f_3\bigg] \;, \\
\Pi^R (x, \mathbf{k}) &=& \int \frac{d^3 p_1 d^3 p_2}{(2 \pi)^3}
\frac{1}{\varepsilon_{\mathbf{k}} + \varepsilon_{\mathbf{p}_2} -
\varepsilon_{\mathbf{p}_1} + i \sigma} \delta (\mathbf{k} + \mathbf{p}_2
- \mathbf{p}_1) \bigg[f_1 (1 + f_2) - f_2 (1 + f_1)\bigg] \;, \\
\Pi^K (x, \mathbf{k}) &=& i \int \frac{d^3 p_1 d^3 p_2}{(2 \pi)^2} \delta
(\varepsilon_{\mathbf{k}} + \varepsilon_{\mathbf{p}_2} -
\varepsilon_{\mathbf{p}_1}) \delta (\mathbf{k} + \mathbf{p}_2 -
\mathbf{p}_1) \bigg[f_1 (1 + f_2) + f_2 (1 + f_1) \bigg] \;, 
\end{eqnarray}

where we have used the shorthand $f_i=f(x,\mathbf{p}_i)$ and the details of the computation of the last two terms $\Pi^K(x,\mathbf{k})$ and $\Pi^R(x,\mathbf{k})$ are in the appendix \ref{appPi}. Action \eqref{sefftotal} is the main result of this subsection and corresponds to \emph{stochastic and dissipative/collisional} dynamics for the slow, coherent fields,  arising from fluctuations of the fast fields that have been integrated out.   

\subsection{Step IV: Fast mode (non-coherent) dynamics from the propagator relations}

\begin{figure}[t]
\centering
\includegraphics[scale=1]{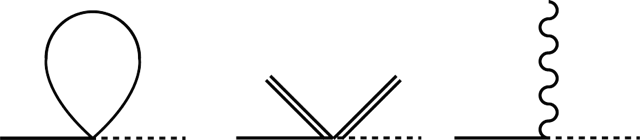}
\caption{Non-coherent self-energy $\Sigma_{0(nc)}^{R}$ at order one in the perturbative parameters, describing the corrections to the propagator $G_{(nc)}^R$, symbolized by the continuous line in the left half and dashed in the right half (the opposite case is $G_{(nc)}^A$). The left diagram is the usual loop from the vertex $\varphi^{q \ast} \varphi^{\ast} \varphi \varphi$. The completely continuous line in the loop corresponds to the non-coherent dressed propagator $G_{(nc)}^K$. The diagrams in the center and right are not closed since they are composed of classical slow fields. They come from the vertices $\Phi_0
\Phi_0^{\ast} \varphi \varphi^{q \ast}$ and $V^{cl}\varphi^{q\ast} \varphi$ respectively. The double line represents the $\Phi_0$ fields and the wavy one the $V^{cl}$ field.}
\label{fig:feynman}
\end{figure}

To get the effective action \eqref{sefftotal} we integrated out the non-coherent (fast) part, meaning that its dynamics is not available from the action anymore. In order to get the equations that describe this component we will now focus on the Schwinger-Dyson equation \eqref{SDeq1} and we will use the corresponding Feynman diagrams to construct the self-energies. Multiplying this equation by $G_{0 (nc)}^{-1}$ from the left we get
\begin{equation}
\left(G_{0 (nc)}^{-1}-\Sigma_{(nc)}\right)\otimes G_{(nc)} = I \;, 
\end{equation}
where $I$ is the unit matrix. Explicitly, we have that
\begin{equation}
\label{sdprop}
\left(\begin{array}{cc}
  0 & i \partial_t - H_0 - \Sigma_{0(nc)}^A - \Sigma_{(nc)}^A\\
  i \partial_t - H_0 - \Sigma_{0(nc)}^R - \Sigma_{(nc)}^R & - \Sigma_{(nc)}^K
\end{array}\right) \left(\begin{array}{cc}
  G_{(nc)}^K & G_{(nc)}^R\\
  G_{(nc)}^A & 0
\end{array}\right) = \left(\begin{array}{cc}
  1 & 0\\
  0 & 1
\end{array}\right) \;, 
\end{equation}
where we have split the advanced and retarded self-energies in two pieces. The first one, $\Sigma_{0(nc)}^{R}$ (analogously for $\Sigma_{0(nc)}^{A}$), will be the corrections of order one in the perturbative parameters, which we display in figure \ref{fig:feynman}.
The first element (left diagram in figure \ref{fig:feynman}) is a typical loop coming from the vertex $\varphi^{q \ast} \varphi^{\ast} \varphi \varphi$ and the other two diagrams are not closed in loops since they are formed with classical slow fields. The rule to deal with them is the following: the classical fields don't enter in loops, but they enter as corrections just as they were a source through the expression defined by the corresponding vertex. Thus,
\begin{equation}
\Sigma_{0(nc)}^A(x,x')= \Sigma_{0(nc)}^R(x,x')=\left[m V^{cl}+2g(n_c+\tilde{n})\right] \delta(x - x')=V_{nc}\,\delta(x - x') \;. 
\end{equation}
The second piece of self-energies, without subindex $0$, will be composed by the higher order corrections. 

The matrix equation \eqref{sdprop} gives rise to the following relations:
\begin{eqnarray}
 \int d^4x' \left[(G_{0(nc)}^A)^{- 1}(x,x') - V_{nc} - \Sigma_{(nc)}^A(x,x')\right] G_{(nc)}^A(x',y) = \delta (x - y) \;, \label{eqga}\\
  \int d^4x' \left[(G_{0(nc)}^R)^{-1}(x,x') - V_{nc} - \Sigma_{(nc)}^R(x,x')\right] G_{(nc)}^R(x',y) = \delta (x-y) \;, \label{eqgr}\\
  \int d^4x' \left[(G_{0(nc)}^R)^{- 1}(x,x') - V_{nc} - \Sigma_{(nc)}^R(x,x')\right] G_{(nc)}^K(x',y) = \int d^4x' \Sigma_{(nc)}^K(x,x') G_{(nc)}^A(x',y) \;.  \label{eqgk}
\end{eqnarray}
We use the relation \eqref{gkeldF}, which relates $G_{(nc)}^K$ with the Hermitian function $F$. Furthermore, from equation \eqref{defspartprop} we observed that this function $F$ is related with the occupation number of the particles $f$ as 
\begin{equation}
\label{relationf}
F=2f+1 \;, 
\end{equation}
and with it and multiplying \eqref{eqgk} from the right by $ \left( \left(G_{0(nc)}^A\right)^{- 1} - V_{nc} - \Sigma_{(nc)}^A \right)$, using \eqref{eqga} and \eqref{eqgr} and after some algebra, we arrive at
\begin{equation}
\label{eqpropgk}
  - \left[ i \partial_t + \frac{1}{2 m} \nabla^2 - V_{ext} (y)-V_{nc}(y), F
  \right] = \Sigma_{(nc)}^K - [\Sigma_{(nc)}^R, F] - F (\Sigma_{(nc)}^R - \Sigma_{(nc)}^A) \;. 
\end{equation}
After a Wigner transform in \eqref{eqpropgk}, and using \eqref{relationf}, we get
\begin{eqnarray}
\label{kineticeq}
& & \left( 1 - \frac{\partial}{\partial \varepsilon} \Re (\Sigma_{(nc)}^R)
  \right) \frac{\partial}{\partial t} f(x,\mathbf{p}) + \left( \frac{\mathbf{p}}{m} +
  \nabla_{\mathbf{p}} \Re (\Sigma_{(nc)}^R) \right)\cdot
  \nabla f(x,\mathbf{p}) \nonumber\\
& & - \nabla \bigg(V_{ext}+V_{nc}
  + \Re (\Sigma_{(nc)}^R)\bigg) \cdot \nabla_{\mathbf{p}} f(x,\mathbf{p}) = \frac{1}{2}I^{coll} [f] \;, 
\end{eqnarray}
where
\begin{equation}
\label{collisional}
  I^{coll} [f] = i \Sigma_{(nc)}^K + 2 (2f+1) \Im (\Sigma_{(nc)}^R) \;, 
\end{equation}
and here $x$ corresponds to the central point coordinates $({x+x'})/{2}$, $\mathbf{p}$ is the momentum related with the Fourier transform of the relative coordinate $x-x'$, and $\Re$, $\Im$ respectively denote real and imaginary parts. We have left the explicit computation of the self-energies $\Sigma_{(nc)}^R$ and $\Sigma_{(nc)}^K$ for appendix \ref{apppart}, where we have observed that the real part of $\Sigma_{(nc)}^R$ is zero and as a consequence the non-coherent self-energies only contribute to the collisional terms.

\subsection{Step V: The full equations of the system}\label{Full_Eqs}

Now we can gather all the elements to present the equations for the system and summarise the important equations. In the effective action, as we mentioned before, we will make the redefinition $\Phi_0 \to \sqrt{2}\Phi_0$ to get as normalization of the field the coherent particle number $N_c$. Also, for convenience, we make the redefinition $\xi_1\to\sqrt{2}\xi_1$. After this, we vary with respect to $\Phi^{q\ast}$ and $V^q$ to get the equations of motion of the coherent field and the field $V^{cl}$ respectively. In addition, we use the equation \eqref{kineticeq}, considering the results of the self-energies computed in appendix \ref{apppart}. Thus, our equations are

\begin{eqnarray}
& & i \frac{\partial \Phi_0 (x)}{\partial t} = \left( - \frac{1}{2 m} \nabla^2 +
V_{ext}(x) + V_c(x) \right) \Phi_0 (x)\nonumber\\
& & - i R(x) \Phi_0 (x) + \xi_1(x) - 2 g \int d^4 x' \Pi^R (x', x) V_{nc} (x') \Phi_0 (x) +
g \xi_2(x) \Phi_0(x) \;, \label{condensateeq}\\
& & \frac{1}{4\pi G}\nabla^2 V^{cl} (x) = m (n_c (x) + \tilde{n} (x)) - m
\int d^4 x' \Pi^R (x', x) V_{nc} (x') + \frac{1}{2} m \xi_2(x) \;, \label{vcleq}\\\
& & \frac{\partial f}{\partial t} + \frac{\mathbf{p}}{m} \cdot\nabla f - \nabla \bigg(V_{ext}(x)
+ V_{nc}(x)\bigg) \cdot \nabla_{\mathbf{p}} f =
\frac{1}{2} (I_a + I_b) \;, \label{particleeq}
\end{eqnarray}
where we recall that the mean field potentials for the coherent and non-coherent parts are respectively
\begin{eqnarray}
V_c (x) &=& m V^{cl} (x) + g (n_c (x) + 2 \tilde{n} (x)) \;, \\
V_{nc} (x) &=& m V^{cl} (x) + 2 g (n_c (x) +
\tilde{n} (x)) \;, 
\end{eqnarray}
and the number densities for the coherent and non-coherent part are
\begin{equation}
n_c=|\Phi_0|^2, \quad \tilde{n}=\int \frac{d^3 p}{(2\pi)^3}f(x,\mathbf{p})\,.
\end{equation}
The terms in the right hand side of \eqref{particleeq} corresponding to collisional terms of this Boltzmann equation are given by
\begin{eqnarray}
I_a &=& 4 g^2 n_c \int \frac{d^3 p_1 d^3 p_2 d^3 p_3}{(2 \pi)^2}
\delta (\varepsilon_c + \varepsilon_{\mathbf{p}_1} -
\varepsilon_{\mathbf{p}_2} - \varepsilon_{\mathbf{p}_3}) \delta
(\mathbf{p}_1 - \mathbf{p}_2 - \mathbf{p}_3)\nonumber\\
& & \times (\delta (\mathbf{p}_1 - \mathbf{p}) - \delta (\mathbf{p}_2 -
\mathbf{p}) - \delta (\mathbf{p}_3 - \mathbf{p})) ((1 + f_1) f_2 f_3 -
f_1 (1 + f_2) (1 + f_3)) \;,  \nonumber\\
I_b &=& 4 g^2 \int \frac{d^3 p_2 d^3 p_3 d^3 p_4}{(2 \pi)^5} \delta
(\varepsilon_{\mathbf{p}} + \varepsilon_{\mathbf{p}_2} -
\varepsilon_{\mathbf{p}_3} - \varepsilon_{\mathbf{p}_4}) \delta
(\mathbf{p} + \mathbf{p}_2 - \mathbf{p}_3 - \mathbf{p}_4) \nonumber\\
& & \times [f_3 f_4 (1 + f) (1 + f_2) - f f_2 (1 + f_3) (1 + f_4)] \,. \label{eqcolls}
\end{eqnarray}

From the explicit expression for $R$ in \eqref{rdef} we have that
\begin{eqnarray}
R &=& g^2
\int \frac{d^3 p_1 d^3 p_2 d^3 p_3}{(2 \pi)^5} \delta
(\varepsilon_c + \varepsilon_{\mathbf{p}_1} -
\varepsilon_{\mathbf{p}_2} - \varepsilon_{\mathbf{p}_3}) \delta ( \mathbf{p}_1
- \mathbf{p}_2 - \mathbf{p}_3) \bigg[ f_1 (1 + f_2) (1 + f_3) - (1 + f_1) f_2 f_3 \bigg]\nonumber\\
&=& \frac{1}{4 n_c} \int \frac{d^3 p}{(2 \pi)^3} I_a \,.\label{rdef2again}
\end{eqnarray}

Also,
\begin{equation}
\Pi^R (x, \mathbf{k}) = \int \frac{d^3 p_1 d^3 p_2}{(2 \pi)^3}
\frac{1}{\varepsilon_{\mathbf{k}} + \varepsilon_{\mathbf{p}_2} -
\varepsilon_{\mathbf{p}_1} + i \sigma} \delta (\mathbf{k} + \mathbf{p}_2
- \mathbf{p}_1) \bigg[f_1 (1 + f_2) - f_2 (1 + f_1)\bigg] \;, 
\end{equation}
and the noise terms $\xi_1$ and $\xi_2$ are Gaussian, satisfying the correlations
\begin{eqnarray}
\langle \xi_1^{\ast} (x) \xi_1 (x') \rangle &=& \frac{i}{2} \Sigma_{(c)}^K (x)\delta(x-x') \;, \\
\langle \xi_2 (x) \xi_2 (x') \rangle&=&- 2 i \Pi^K (x, x') \;, \label{realcorrelators1}
\end{eqnarray}
where we remember that
\begin{eqnarray}
\Sigma_{(c)}^K(x) &=& - 2 i g^2
\int \frac{d^3 p_1 d^3 p_2 d^3 p_3}{(2 \pi)^5} \delta
(\varepsilon_c + \varepsilon_{\mathbf{p}_1} -
\varepsilon_{\mathbf{p}_2} - \varepsilon_{\mathbf{p}_3}) \delta (\mathbf{p}_1 - \mathbf{p}_2 - \mathbf{p}_3) \nonumber\\
& & \times \bigg[f_1 (1 + f_2) (1 + f_3) + (1 +
f_1) f_2 f_3\bigg] \;,  \label{sigkdefagain}\\
\Pi^K (x, \mathbf{k}) &=&  i \int \frac{d^3 p_1 d^3 p_2}{(2 \pi)^2} \delta
(\varepsilon_{\mathbf{k}} + \varepsilon_{\mathbf{p}_2} -
\varepsilon_{\mathbf{p}_1}) \delta (\mathbf{k} + \mathbf{p}_2 -
\mathbf{p}_1) \bigg[f_1 (1 + f_2) + f_2 (1 + f_1)\bigg]\,. \label{pikdefagain}
\end{eqnarray}

In order to obtain a generalization in the expanding universe, we make the following transformations: we pass from physical coordinates to comoving ones $\mathbf{r}\to a \mathbf{r}$, which means that we transform the derivatives $\nabla\to {a}^{-1}\nabla$. Also, we perform the change $\mathbf{p}\to a^{-1}\mathbf{p}$ and the densities are transformed to the comoving ones $n_c\to a^{-3}n_c$ and $\tilde{n}\to a^{-3}\tilde{n}$. The fields appearing in the equations are then to be understood as comoving, relating to the physical ones as  $\Phi_0=\Phi^{\rm phys}_0 \,a^{-3/2}$ and $f =  f^{\rm phys} a^{-3}$, consistent with the definitions used in \cite{Proukakis:2023nmm}. The resulting equations are those presented in section \ref{Equations} and constitute the main equations for a bosonic system composed by slow (coherent) and fast (non-coherent) particles, interacting with an additional potential that we have encoded in $V^{cl}$ in an expanding universe.

The set of equations generalises the result obtained in \cite{Proukakis:2023nmm}, where it was shown that in the limit $g\to 0$ and when all modes are coherent, we recover the commonly used Fuzzy Dark Matter equations, and when all mades are incoherent, we recover the typical Vlasov-Poisson equations which are used to describe CDM. Also, if we had worked only up to order one in the quantum components $V^q$ and $\Phi^q$ and turning off the gravitational interaction we would have recovered the ZNG models \cite{zaremba1999dynamics,griffin_nikuni_zaremba_2009}, used in Cold Atom physics to describe bosonic atoms. The incorporation up to second order in quantum components allowed us to get the noises $\xi_1$ and $\xi_2$, which can relate with the Stochastic Projected Gross-Pitaevskii equations~\cite{Gardiner_2003,bradley2006stochasticgrosspitaevskiiequationiii,PhysRevA.77.033616,Blakie-AdvPhys-2008} 
in the absence of gravity, when we additionally ignore the Boltzmann equation for the incoherent particles and we consider them as described by an equilibrium Rayleigh-Jeans distribution, as a semi-classical approximation to the full Bose-Einstein distribution.
Alternatively, ignoring terms $\Pi^R$, associated noise $\xi_2$ and gravity, our equations coincide with the alternative cold atom stochastic formulation of Stoof~\cite{stoof1999coherent,Duine_2001} and its resulting near-equilibrium stochastic equation~\cite{stoof_dynamics_2001}. The relation between such models in the cold atom context is discussed in our companion paper~\cite{Proukakis-coldatom-stochastic}.


\section{Thermal state and fluctuation-dissipation relation}\label{Thermal-plus}

In the equations we derived above, the fast (incoherent) modes are described via an N-body phase space distribution function $f$ which is to be solved in conjunction with the wave-like equation for the slow (coherent) component - see \ref{Full_Eqs}. The equations involve dissipative/collisional terms and stochastic forces where a characteristic pattern with similar factors of $f$ appears between the  kernels and the variances of the corresponding stochastic forces. As a consistency check, we now confirm that when thermal equilibrium is assumed for the fast modes described by $f$, standard fluctuation-dissipation relations are recovered.

Indeed, if we take the equilibrium Bose-Einstein distribution for the fast modes
\begin{equation}
f(\varepsilon_i,\mu) = \frac{1}{e^{\beta (\varepsilon_i - \mu)} - 1}
\end{equation}
where $\beta=1/k_B T$ and  $\epsilon_i$ is the energy of the particle $i$, and we apply this distribution in the equations \eqref{sigkdefagain} and \eqref{pikdefagain}, we can show that
\begin{eqnarray}
i R (x) &=& - \frac{1}{2} \Sigma_{(c)}^K (x) \frac{1}{1 + 2 f (\varepsilon_c,\mu)} \label{equilibriumSk}\\
\Pi^R (x, \mathbf{p}) - \Pi^A (x, \mathbf{p})&=& \Pi^K (x, \mathbf{p}) \frac{1}{1 + 2f(\varepsilon_\mathbf{p},0)} \label{equilibriumPk}
\end{eqnarray}
The function $\Pi^A(x,\mathbf{p})$ is obtained from $\Pi^R(x,\mathbf{p})$ knowing that $\Pi^A(x,x')=\Pi^R(x',x)$ and proceeding as in Appendix \ref{appPi}. Then, we use that $G^R(p)-G^A(p)=-2\pi i \delta(\varepsilon + \varepsilon_c -\varepsilon_\mathbf{p})$ to arrive to \eqref{equilibriumPk}. In the above expressions $\varepsilon_c$ 
and $\varepsilon_\mathbf{k}$ 
are defined in \eqref{eq:epsilon_c} and \eqref{eq:epsilon_k} respectively 
as the energies of particles in the coherent and incoherent bands respectively. 
Notice that since we have used the Bose-Einstein distribution, we have recovered in our equation \eqref{equilibriumSk} the results obtained in~\cite{stoof1999coherent,Duine_2001}  in the context of the stochastic Gross-Pitaevskii equation.

We can simplify our expressions by the use of the function $F$, defined as $F=2f+1$ in \eqref{defspartprop}, which for the Bose-Einstein distribution reads:
\begin{equation}
F (\varepsilon, \mu) = \coth \left( \frac{\beta}{2} (\varepsilon - \mu)
\right)
\end{equation}
Also, we use that $\Sigma^R_{(c)} (x) = - i R$, $\Sigma_{(c)}^A=\Sigma_{(c)}^{R\ast}$ (see Appendix \ref{appWig}). 
With that, equations \eqref{equilibriumSk} and \eqref{equilibriumPk} become
\begin{eqnarray}
\Sigma_{(c)}^K (x) &=& F (\varepsilon_{c}, \mu) (\Sigma^R_{(c)} (x) -
\Sigma^A_{(c)} (x)) \\
\Pi^K (x, \mathbf{p}) &=& F (\varepsilon_{\mathbf{p}}, 0) (\Pi^R (x, \mathbf{p}) -
\Pi^A (x, \mathbf{p}))\,,
\end{eqnarray}
recovering the standard fluctuation-dissipation theorem for bosonic particles \cite{kamenev_2011}.

\section{Final remarks}\label{Final}

In this work we have extended our earlier  first-principles derivation \cite{Proukakis:2023nmm} of a set of dynamical equations for a generic dark matter model involving a scalar particle with mass $m$ and quartic self-interaction with self-coupling strength $g$. Starting from the non-relativistic, non-equilibrium partition function, we describe the system in terms of two components, one involving fast modes described by a kinetic equation for their phase space distribution $f$, also dubbed the non-coherent part here, and one involving slower modes with longer de-Broglie wavelengths, which are described by a wave equation, all coupled via gravity and their local self-interaction. We quantified the distinction between coherent and incoherent components by comparing the slowness of the coherent field $\Phi_0(x)$ to the rapidity with which the two-point function of the fast fields $G^K(x,\Delta x)$ goes to zero with increasing $\Delta x$. The fast-slow separation could then be drawn within an energy range, provided such a separation still makes sense. An absolute lower limit for demarcating this fast-slow separation would be the energy of the fully condensed Penrose-Onsager mode, with an upper limit set by the regime of modes with low occupation numbers which could not be part of $\Phi_0$.   

The new feature of our work here and the resulting equations is the derivation of a set of two dissipation/collisional terms and two corresponding stochastic noise terms that affect the coherent part's dynamics. These terms reflect the influence of the fluctuations of the faster modes and arise when these faster modes are integrated out to provide effective equations for the slow modes. The kinetic equation for $f$ is in turn obtained from the  Schwinger-Dyson Equations of the fast modes.  The new dissipation/collisional kernels and the variances of the stochastic noise forces are expressed in terms of the fast modes' phase space distribution and come in two pairs, depending on the number of $f$ factors appearing in the relevant expressions. Their forms ensure that the expected Fluctuation-Dissipation relations are satisfied if the fast particles are taken to obey the thermal Bose-Einstein distribution, confirming the consistency of our computation. Note that while the $\xi_2$ noise is multiplicative, being proportional to the coherent component, the $\xi_1$ noise is additive. Hence, the fast fluctuations described by the latter can source the growth of $\Phi_0$ ab initio. Interestingly, $\xi_2$ and the corresponding dissipation kernel also affect the gravitational potential via additions to the Poisson equation. 

All the new terms discussed here are proportional to the self-coupling $g$ and hence the fluctuations we capture with our present approximations do not involve those of the gravitational sector but are restricted to fluctuations whose effects are mediated by the self-coupling. Gravitational fluctuation effects would require us to go to higher order in $V^q$ and, presumably, $\Phi^q$, a task we leave for future investigations. It would nevertheless be interesting to see how our dynamical equations may be used to study the soliton nucleation process of scalar dark matter \cite{Levkov:2018kau, Chen:2021oot, Kirkpatrick:2021wwz}. It would also be interesting to compare our dynamical equations with other kinetic approaches \cite{Bar-Or:2018pxz, Chavanis:2020upb}, kinetic relaxation \cite{Jain:2023ojg,Jain:2023tsr} or the non-equilibrium QFT formalism as in \cite{Friedrich:2019zic}, which however does not use self-interactions, or in \cite{Ai:2023qnr} which utilises the 2PI effective action.

\section*{Acknowledgements}
\noindent This work is supported by the Leverhulme Trust, Grant no.
RPG-2021-010. 

\appendix

\section{Wigner transforms of the coherent self-energies in the effective action}\label{appWig}

We observe that the terms containing the coherent self-energies $\Sigma_{(c)}^R$,$\Sigma_{(c)}^A$ and $\Sigma_{(c)}^K$ in \eqref{effaction0b} have a double integration ($d^4x d^4x'$). However, it is possible to reduce the double integration to a single one using a Wigner transform. We will analyse in detail the term containing the retarded self-energy for the coherent (slow) part. The same steps can be used for the advanced and Keldysh self-energies, so we will pay more attention to this first case, briefly sketching the results for the other self-energies.

\subsection{Coherent Retarded Self-Energy $\Sigma_{(c)}^R$ and $R$} 
The term containing the retarded self-energy for the coherent part is
\begin{eqnarray}
\label{actsigmar}
\int d^4 x \int d^4 x' \Phi^{q \ast} (x') (- \Sigma_{(c)}^R (x, x'))
\Phi_0 (x) &=& \frac{g^2}{2} \int d^4 x \int d^4 x' \Phi^{q \ast} (x')
\bigg( 2 G_{(nc)}^K (x', x) G_{(nc)}^R (x, x') G_{(nc)}^K (x, x') \nonumber\\
& & + G_{(nc)}^A \left( x', x \right) (G_{(nc)}^K
(x, x'))^2 + G_{(nc)}^A (x', x) (G_{(nc)}^R (x, x'))^2 \bigg) \Phi_0 (x) \;. \nonumber\\
\end{eqnarray}
For later convenience, we can replace the elements $G_{(nc)}^R (x, x')$ in the last term inside the parenthesis in \eqref{actsigmar} by $(G_{(nc)}^R (x, x') - G_{(nc)}^A (x, x'))$,
since $G_{(nc)}^A (x', x) G_{(nc)}^A (x, x') = 0$. With this, we have that
\begin{eqnarray}
\label{sigcondrtmp}
\int d^4 x \int d^4 x' \Phi^{q \ast} (x') (- \Sigma_{(c)}^R (x, x'))
\Phi_0 (x)&=&
g^2 \int d^4 x \int d^4 x' \Phi^{q \ast} (x') G_{(nc)}^K (x', x) G_{(nc)}^R (x, x')
G_{(nc)}^K (x, x') \Phi_0 (x) \nonumber\\
& & + \frac{g^2}{2} \int d^4 x \int d^4 x' \Phi^{q \ast} (x') G_{(nc)}^A (x', x) G_{(nc)}^K
(x, x') G_{(nc)}^K (x, x') \Phi_0 (x) \nonumber\\
& & + \frac{g^2}{2} \int d^4 x \int d^4 x' \Phi^{q \ast} (x') G_{(nc)}^A (x', x) (G_{(nc)}^R
(x, x') - G_{(nc)}^A (x, x')) \nonumber\\
& & \ \times(G_{(nc)}^R (x, x') - G_{(nc)}^A (x, x')) \Phi_0 (x) \;. 
\end{eqnarray}
We will only look in detail at the procedure for the integral in the first line in \eqref{sigcondrtmp}, which is
\begin{equation}
J=\int d^4 x \int d^4 x' \Phi^{q \ast} (x') G_{(nc)}^K (x', x) G_{(nc)}^R (x, x')
G_{(nc)}^K (x, x') \Phi_0 (x) \;, 
\end{equation}
and we will repeat the same procedure for the next lines, skipping for brevity the details. In this term we use the Wigner transform 
\begin{equation}
\label{wignertra}
G (x, x') =
\underset{p}{\sum} e^{i p (x - x')} G \left( \frac{x + x'}{2}, p \right) \;, 
\end{equation}
where $G$ corresponds to any function of two points $x$ and $x'$ (remembering that $x$ and $x'$ are space-time coordinates) and the sum $\underset{p}{\sum}=\underset{\mathbf{p}}{\sum}\int \frac{d\epsilon}{2\pi}$, with $\epsilon$ the energy and $\mathbf{p}$ the corresponding momentum. Therefore,
\begin{eqnarray}
J &=& \int d^4 x \int d^4 x' \Phi^{q \ast} (x') \underset{p_1, p_2, p_3}{\sum}
e^{i (p_2 + p_3 - p_1) (x - x')} G_{(nc)}^K \left( \frac{x + x'}{2},
p_1 \right) G_{(nc)}^R \left( \frac{x + x'}{2}, p_3 \right) \nonumber\\
& & \times G_{(nc)}^K \left( \frac{x + x'}{2}, p_2 \right) \Phi_0 (x) \;. 
\end{eqnarray}

%
%

Now, we make the change of variables
\begin{equation}\label{eq:Keldysh_coords}
x_a = \frac{x + x'}{2}, \qquad x_b = x - x' \;, 
\end{equation}
where we observe that $x_a$ corresponds to the central point coordinate and $x_b$ is the relative coordinate. With these new variables:
\begin{eqnarray}
J &=& \int d^4 x_a \int d^4 x_b \Phi^{q \ast} \left( x_a - \frac{x_b}{2}
\right) \underset{p_1, p_2, p_3}{\sum} e^{i (p_2 + p_3 - p_1) x_b}
G_{(nc)}^K (x_a, p_1) G_{(nc)}^R (x_a, p_3) \nonumber\\
& & \times G_{(nc)}^K
(x_a, p_2) \Phi_0 \left( x_a + \frac{x_b}{2} \right) \;. 
\end{eqnarray}
Considering that $\Phi_0$ is a slow quantity - see discussion in section \ref{sec:cut} and in particular eqn.~\eqref{eq:gradient_bound} - we can expand around $x_a$ to write, to leading order
\begin{eqnarray}
J &\simeq& \int d^4 x_a \int d^4 x_b \Phi^{q \ast} (x_a) \underset{p_1, p_2,
p_3}{\sum} e^{i (p_2 + p_3 - p_1) x_b} G_{(nc)}^K (x_a, p_1)
G_{(nc)}^R (x_a, p_3) \nonumber\\
& & \times G_{(nc)}^K (x_a, p_2) \Phi_0 (x_a) \;. \label{eq:local_approx} 
\end{eqnarray}
Doing the integration in $x_b$ and renaming $x_a \rightarrow x$ we have that
\begin{equation}
J \simeq \int d^4 x \, \Phi^{q \ast} (x) \left[ (2 \pi)^4 \underset{p_1, p_2,
p_3}{\sum} \delta (p_2 + p_3 - p_1) G_{(nc)}^K (x, p_1)
G_{(nc)}^R (x, p_3) G_{(nc)}^K (x, p_2) \right] \Phi_0 (x) \;. 
\end{equation}
%
%
%

Repeating the same steps for the next two terms in \eqref{sigcondrtmp} we end up with
\begin{equation}
\int d^4 x \int d^4 x' \, \Phi^{q \ast} (x') (- \Sigma_{(c)}^R (x, x'))
\Phi_0 (x) \simeq - \int d^4 x \, \Phi^{q \ast} (x) \Sigma^R_{(c)} (x) \Phi_0 (x) \;, 
\end{equation}
where
\begin{eqnarray}
\Sigma_{(c)}^R(x) &=& - (2 \pi)^4 \frac{g^2}{2} \underset{p_1, p_2, p_3}{\sum}
\delta (p_2 + p_3 - p_1) \bigg[ 2 G_{(nc)}^K (x,p_1) G_{(nc)}^R
(x,p_3) G_{(nc)}^K (x,p_2) \nonumber\\
& & + G_{(nc)}^A (x,p_1) G_{(nc)}^K
(x,p_3) G_{(nc)}^K ( x,p_2) + G_{(nc)}^A (x,p_1)
(G_{(nc)}^R (x,p_2) \nonumber\\
& & - G_{(nc)}^A (x,p_2)) (G_{(nc)}^R (x,p_3) -
G_{(nc)}^A (x,p_3)) \bigg] \;. 
\end{eqnarray}
%
%
Now, we use \cite{kamenev_2011} 
\begin{equation}
\label{keldrelations}
(G_{(nc)}^R (p) - G_{(nc)}^A (p)) = - 2 \pi i \, \delta (\varepsilon +
\varepsilon_c - \varepsilon_{\mathbf{p}}), \qquad G_{(nc)}^K (p) = - 2 \pi i
\tilde{F}(p) \, \delta (\varepsilon +
\varepsilon_c - \varepsilon_{\mathbf{p}}) \;, 
\end{equation}
where the retarded and advanced propagators are given by
\begin{equation}
\label{retandandvprop}
G_{(nc)}^R (p) = \frac{1}{\varepsilon + \varepsilon_c - \varepsilon_{\mathbf{p}} + i \sigma}, \quad
G_{(nc)}^A (p) = \frac{1}{\varepsilon + \varepsilon_c - \varepsilon_{\mathbf{p}} - i \sigma} \;, 
\end{equation}
which contains the $i \sigma$ prescription to ensure the correct causal structure, and $\varepsilon_c$ corresponds to the coherent component energy, given by \eqref{eq:epsilon_c}. Here we note that $\varepsilon_c$ appears in the non-coherent propagators because the energy of the non-coherent modes by definition lies above the energy of the coherent modes - see discussion in section \ref{sec:cut} and equation \eqref{eq:energy_hierarchy}. Consequently, $i \partial_t \varphi \to \left(\varepsilon + \varepsilon_c\right)\varphi$ and hence the appearance of $\varepsilon+\varepsilon_c$ in the propagators.

We now symmetrize the first term of $\Sigma_{(c)}^R(x)$ between $p_2$ and $p_3$. Using these relations, we split $\Sigma_{(c)}^R$ in a real and imaginary part and observe that every term in the real part contains $(G^R + G^A)$, which will be zero after the energy integration. Therefore, we are left with just the imaginary part. Then, we do the energy integration in the imaginary part and using that $F=2f+1$ we get that
\begin{eqnarray}
\label{sigmacondRfin}
\Im (\Sigma_{(c)}^R) &=& - 2\pi g^2 \underset{\mathbf{p}_1,
\mathbf{p}_2, \mathbf{p}_3}{\sum} \delta (\varepsilon_c +
\varepsilon_{\mathbf{p}_1} - \varepsilon_{\mathbf{p}_2} -
\varepsilon_{\mathbf{p}_3}) \, \delta(\mathbf{p}_1-\mathbf{p}_2-\mathbf{p}_3) \nonumber\\
& & \times\bigg[f_1 (1 + f_2) (1 + f_3) - (1 + f_1) f_2 f_3\bigg] \;, 
\end{eqnarray}
where we used the shorthand $f_i=f(x,\mathbf{p}_i)$. Considering a thermodynamical limit, we turn the sum on the momenta to an integral and get that
\begin{eqnarray}
\label{erre}
R &=& - \Im (\Sigma_{(c)}^{R}) = g^2
\int \frac{d^3p_1 d^3p_2 d^3p_3}{(2\pi)^5} \, \delta
(\varepsilon_c + \varepsilon_{\mathbf{p}_1} -
\varepsilon_{\mathbf{p}_2} - \varepsilon_{\mathbf{p}_3}) \, \delta (\mathbf{p}_1
- \mathbf{p}_2 - \mathbf{p}_3) \nonumber\\
& & \times \bigg[ f_1 (1 + f_2) (1 + f_3) - (1 + f_1) f_2 f_3 \bigg] \;, 
\end{eqnarray}

and with that
\begin{equation}
\int d^4 x \int d^4 x' \Phi^{q \ast} (x) (- \Sigma_{(c)}^R (x, x'))
\Phi_0 (x') \simeq \int d^4 x \Phi^{q \ast} (x) i R \Phi_0 (x) \;. 
\end{equation}
Analogously we can do the same for the term involving $\Sigma_{(c)}^A (x, x')$ and get
\begin{equation}
\int d^4 x \int d^4 x' \Phi_0^{\ast} (x) (- \Sigma_{(c)}^A (x, x'))
\Phi^q (x') \simeq  - \int d^4 x \Phi_0^{\ast} (x) i R \Phi^q (x) \;. 
\end{equation}
Notice that this approximate result preserves the relation $\Sigma_{(c)}^A=\Sigma_{(c)}^{R\ast}$.

\subsection{Coherent Keldysh Self-Energy $\Sigma^K_{(c)}$}
The Keldysh self-energy is given by
\begin{eqnarray}
- \int d^4 x \int d^4 x' \Phi^{q \ast} (x') \Sigma_{(c)}^K (x, x')
\Phi^q (x) &=& \frac{g^2}{2} \int d^4 x \int d^4 x' \Phi^{q \ast} (x') \bigg( G_{(nc)}^K
(x', x) (G_{(nc)}^K (x, x'))^2 + G_{(nc)}^K (x', x) (G_{(nc)}^R (x, x'))^2 \nonumber\\
& & + G_{(nc)}^K (x', x) \left( G_{(nc)}^A
\left( x, x' \right) \right)^2 + 2 G_{(nc)}^A (x', x) G_{(nc)}^K (x, x') G_{(nc)}^R (x, x') \nonumber\\
& & + 2 G_{(nc)}^R
(x', x) G_{(nc)}^A (x, x') G_{(nc)}^K (x, x') \bigg) \Phi^q (x) \;. 
\end{eqnarray}
After using the identities
\begin{eqnarray}
(G_{(nc)}^R (x, x') - G_{(nc)}^A (x, x')) (G_{(nc)}^R (x, x') - G_{(nc)}^A (x, x')) &=& (G_{(nc)}^R (x, x'))^2 +
(G_{(nc)}^A (x, x'))^2 \;, \\
(G_{(nc)}^R (x', x) - G_{(nc)}^A (x', x)) (G_{(nc)}^R (x, x') - G_{(nc)}^A (x, x')) &=& - G^A (x', x) G_{(nc)}^R
(x, x') - G_{(nc)}^R (x', x) G_{(nc)}^A (x, x') \;, \nonumber \\
\end{eqnarray}
performing the Wigner transforms as we did for $\Sigma_{({c})}^R$ and changing variables as in \eqref{eq:Keldysh_coords}, we arrive at
\begin{equation}
\int d^4 x \int d^4 x' \, \Phi^{q \ast} (x') (- \Sigma_{(c)}^K (x, x'))
\Phi^q (x) \simeq - \int d^4 x \, \Phi^{q \ast} (x) \Sigma^K_{(c)} (x) \Phi^q (x) \;, 
\end{equation}

where
\begin{eqnarray}
\Sigma_{(c)}^K (x) &=& -(2\pi)^4 \frac{g^2}{2} \underset{p_1,
p_2, p_3}{\sum} \bigg[ G_{(nc)}^K (x, p_1) G_{(nc)}^K (x, p_2) G_{(nc)}^K (x, p_3) \nonumber\\
& & + G_{(nc)}^K (x, p_1) (G_{(nc)}^R (x,
p_2) - G_{(nc)}^A (x, p_2)) \left( G_{(nc)}^R ( x, p_3 ) - G_{(nc)}^A (x, p_3)
\right) \nonumber\\
& & - 2 G_{(nc)}^K (x, p_3) (G_{(nc)}^R (x, p_1) - G_{(nc)}^A (x, p_1)) (G_{(nc)}^R (x, p_2) - G_{(nc)}^A (x, p_2)) \bigg] \;. 
\end{eqnarray}
Symmetrizing between $p_2$ and $p_3$ in the last term, using the relations in equation \eqref{keldrelations} and the relation $F=2f+1$, doing the energy integration and passing the sum on the momenta to integrals as before, we end up with
\begin{eqnarray}
\Sigma_{(c)}^K(x) &=& -2 i g^2
\int \frac{d^3p_1 d^3p_2 d^3p_3}{(2\pi)^5} \, \delta
(\varepsilon_c + \varepsilon_{\mathbf{p}_1} -
\varepsilon_{\mathbf{p}_2} - \varepsilon_{\mathbf{p}_3}) \, \delta (\mathbf{p}_1 - \mathbf{p}_2 - \mathbf{p}_3) \nonumber\\
& & \times\bigg[f_1 (1 + f_2) (1 + f_3) + (1 +
f_1) f_2 f_3 \bigg] \;. 
\end{eqnarray}

\section{Wigner transforms of Self-energies $\Pi^K(x',x)$ and $\Pi^R(x',x)$}\label{appPi}

\subsection{Term $\Pi^K(x',x)$}

We recall our definition of $\Pi^K(x',x)$ given in \eqref{PiK0} by
\begin{equation}
\Pi^K (x, x') = - \frac{i}{2} [G_{(nc)}^K (x', x) G_{(nc)}^K (x, x') + G^A (x', x) G_{(nc)}^R (x, x') + G_{(nc)}^R (x', x) G_{(nc)}^A (x, x')] \;, 
\end{equation}
and we use the identity
%
%
\begin{equation}
- (G_{(nc)}^R (x, x') - G_{(nc)}^A (x, x')) (G_{(nc)}^R (x', x) - G_{(nc)}^A (x', x)) = G_{(nc)}^A (x, x') G_{(nc)}^R
(x', x) + G_{(nc)}^R (x, x') G_{(nc)}^A (x', x) \;, 
\end{equation}
together with the fact that the Wigner transform is:
\begin{equation}
\label{wignerdef}
\Pi^K (x, k) = \int d x' e^{- i k x'} \Pi^K \left( x + \frac{x'}{2}, x -
\frac{x'}{2} \right) \;, 
\end{equation}
and \eqref{wignertra} to write

\begin{eqnarray}
\Pi^K (x, k) &=& - \frac{i}{2} \int d x' \underset{p_1, p_2}{\sum} e^{i (p_1 -
p_2 - k) x'} \bigg[G_{(nc)}^K (x, p_2) G_{(nc)}^K (x, p_1) \nonumber\\
& & - (G_{(nc)}^R (x, p_2) - G_{(nc)}^A (x, p_2)) (G_{(nc)}^R
(x, p_1) - G_{(nc)}^A (x, p_1)) \bigg] \;. 
\end{eqnarray}
Integrating in $x'$, using \eqref{keldrelations}, doing the energy integral, using $F=2f+1$ and passing the sum on the momenta to integrals, we obtain after some arrangements
\begin{equation}
\Pi^K (x, k) = i \int \frac{d^3 \mathbf{p}_1 d^3 \mathbf{p}_2}{(2 \pi)^2}
\,\delta (\mathbf{p}_1 - \mathbf{p}_2 - \mathbf{k}) 
\,\delta
(\varepsilon_{\mathbf{p}_1} - \varepsilon_{\mathbf{p}_2} -
\varepsilon_{\mathbf{k}}) [f_1 (1 + f_2) + (1 + f_1) f_2] \;. 
\end{equation}

\subsection{Term $\Pi^R(x',x)$}

We now focus on the term $\Pi^R(x',x)$ defined in \eqref{PiR0} by 
\begin{equation}
\Pi^R (x', x) = - \frac{i}{2} \bigg(G_{(nc)}^R (x', x) G_{(nc)}^K (x, x') + G_{(nc)}^K (x', x) G_{(nc)}^A (x,
x') \bigg) \;. 
\end{equation}
Using as before \eqref{wignerdef}, we write
\begin{equation}
\Pi^R (x', x) = - \frac{i}{2} \int d x' \underset{p_1, p_2}{\sum} e^{i(p_1-p_2-k)x'}\bigg(G_{(nc)}^R (x,p_1) G_{(nc)}^K (x,p_2) + G_{(nc)}^K (x,p_1) G_{(nc)}^A (x,
p_2) \bigg) \;. 
\end{equation}

Then, we use the equation \eqref{keldrelations} for $G_{(nc)}^K$, the explicit relations for the retarded and advanced propagators \eqref{retandandvprop} 
and the relation $F=2f+1$, doing the energy integration and passing the sum on the momenta to integrals we get finally that

\begin{equation}
\Pi^R (x, \mathbf{k}) = \int \frac{d^3 p_1 d^3 p_2}{(2 \pi)^3}
\frac{1}{\varepsilon_{\mathbf{k}} + \varepsilon_{\mathbf{p}_2} -
\varepsilon_{\mathbf{p}_1} + i \sigma} \delta (\mathbf{k} + \mathbf{p}_2
- \mathbf{p}_1) \bigg[f_1 (1 + f_2) - f_2 (1 + f_1) \bigg] \;. 
\end{equation}

\section{Non-coherent Particle self-energies computation and collisional terms}\label{apppart}
\begin{figure}[t]
    \centering   \includegraphics[width=1.2\linewidth,]{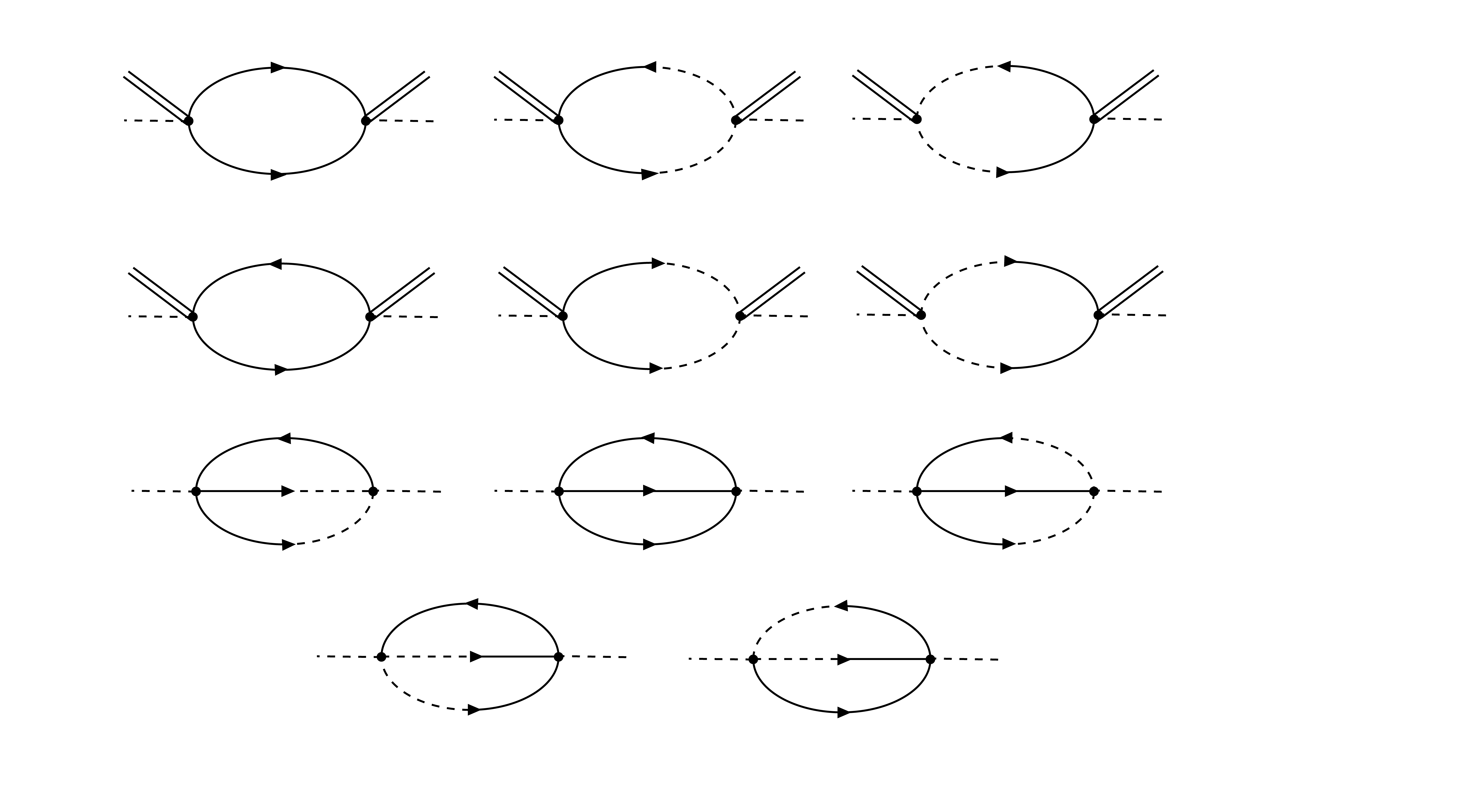}
    \caption{Feynman diagrams for the particle Keldysh self-energy at order $g^2$. The six upper diagrams correspond to $\Sigma^{K}_{(a)}$ (eq. \eqref{sigmaaK}), the remaining corresponds to $\Sigma^{K}_{(b)}$ (eq. \eqref{sigmabK}).
    }
    \label{fig:sigmak}
\end{figure}

The interaction action $S_I$ in equation \eqref{Si}, defines through the terms in equations \eqref{Si2}, \eqref{Si3}, \eqref{Si4} and \eqref{Sia2} the elements for constructing the Feynman diagrams making up the self-energies for the non-coherent particles (see figures \ref{fig:sigmarpart} and \ref{fig:sigmak}). We will compute the retarded and Keldysh self-energies. The advanced one is obtained from the retarded one via complex conjugation.

\subsection{Retarded Non-coherent Particle Self-energy $\Sigma_{(nc)}^R$}

\begin{figure}[t]
    \centering
    \includegraphics[width=1\linewidth,]{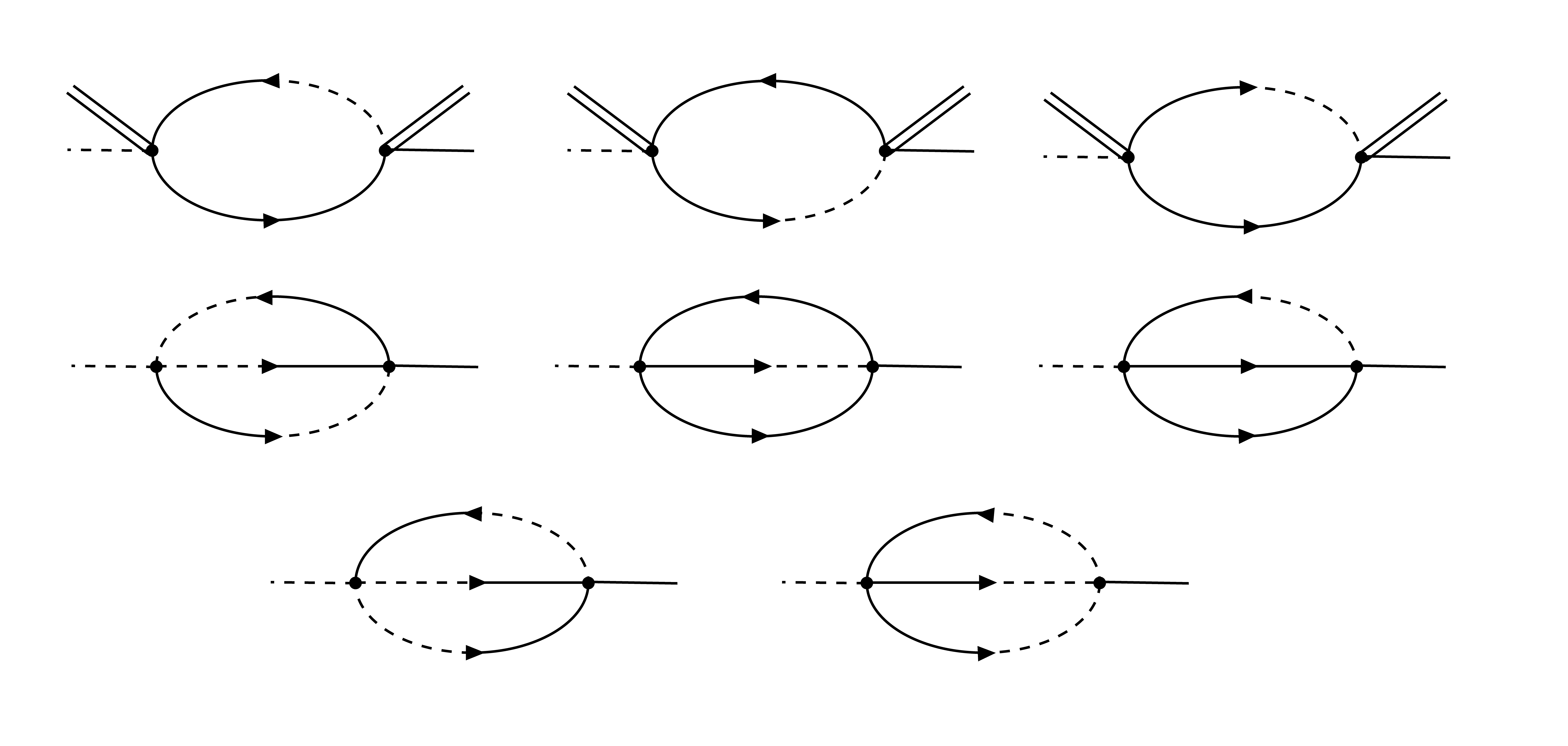}
    \caption{Feynman diagrams for the particle retarded self-energy at order $g^2$. The three upper diagrams correspond to $\Sigma^{R}_{(a)}$ (eq. \eqref{sigmaaR}), the remaining corresponds to $\Sigma^{R}_{(b)}$ (eq. \eqref{sigmabR}). The double lines in the upper diagrams give origin to the $|\Phi_0|^2$ term in eq. \eqref{sigmaaR}. 
    }
    \label{fig:sigmarpart}
\end{figure}

The contributions to the retarded self-energy  $\Sigma_{(nc)}^R$, depicted in figure \ref{fig:sigmarpart} are written as
\begin{equation}
- i \Sigma_{(nc)}^R = - i \Sigma_{(a)}^R  - i \Sigma_{(b)}^R \;, 
\end{equation}
where we have split it in two different pieces 
\begin{eqnarray}
- i \Sigma_{(a)}^R &=& \frac{g^2}{4}  \Phi_0(x') \bigg[2 G_{(nc)}^A (x', x) G_{(nc)}^K (x, x') +
2 G_{(nc)}^R (x, x') G_{(nc)}^K (x', x) \nonumber\\
& & + 2 G_{(nc)}^R (x, x') G_{(nc)}^K (x, x')\bigg]\Phi^\ast_0(x) \;, \label{sigmaaR}\\
- i \Sigma_{(b)}^R &=& i \frac{g^2}{8} \bigg[ 2 G_{(nc)}^R (x', x) G_{(nc)}^A (x, x') G_{(nc)}^R (x,
x') + 4 G_{(nc)}^K (x', x) G_{(nc)}^K (x, x') G_{(nc)}^R (x, x') \nonumber\\
& & + 2 G_{(nc)}^A (x', x) G_{(nc)}^K (x, x') G_{(nc)}^K
(x, x') + 2 G_{(nc)}^A (x', x) G_{(nc)}^A (x, x') G_{(nc)}^A (x, x') \nonumber\\
& & + 2 G_{(nc)}^A (x', x) G_{(nc)}^R (x, x') G_{(nc)}^R (x, x') \bigg] \;. \label{sigmabR}
\end{eqnarray}
Assuming $\Phi_0$ is slow and using the identities
\begin{eqnarray}
G^A (x, x')^2 + G^R (x, x')^2 &=& (G^R (x, x') - G^A (x, x'))^2 \;, \\
G^R (x', x) G^R (x, x') &=& 0 \;, 
\end{eqnarray}
we symmetrize the last term of $\Sigma_{(a)}^R$ in \eqref{sigmaaR} and the second one of $\Sigma_{(b)}^R$ in \eqref{sigmabR}, we take a Wigner transform, similarly to the coherent part in appendix \ref{appWig} and we use the equation \eqref{keldrelations}. Doing this, as we did in the coherent retarded self-energy, we can split in a real and imaginary part. Every term in the real part contains $G^A + G^R$, which becomes zero after the energy integration. Therefore, only the imaginary part survives and we proceed with it doing the energy integrations, using $F=2f+1$ and the fact that in the Schwinger-Keldysh $|\Phi_0|^2=2n_c$. Thus, we get that the imaginary part of the two pieces reads
\begin{eqnarray}
\Im (\Sigma_{(a)}^R) &=& 2 \pi g^2 n_c
\underset{\mathbf{p}}{\sum} \bigg[2 \delta
(\varepsilon_c + \varepsilon_{\mathbf{p}} -
\varepsilon_{\mathbf{k}} - \varepsilon_{\mathbf{p} -
\mathbf{k}}) (f_p - f_{p - k}) \nonumber\\
& & - \delta (\varepsilon_c +
\varepsilon_{\mathbf{k}} - \varepsilon_{\mathbf{p}} -
\varepsilon_{\mathbf{k} - \mathbf{p}}) (f_p + f_{k - p}
+ 1)\bigg] \;, \\
\Im (\Sigma_{(b)}^R) &=& 4 \pi g^2 \underset{\mathbf{p},
\mathbf{q}}{\sum} \delta (\varepsilon_{\mathbf{p}} +
\varepsilon_{\mathbf{k} - \mathbf{q}} - \varepsilon_{\mathbf{k}} -
\varepsilon_{\mathbf{p} - \mathbf{q}}) [f_{k - q} f_p - f_{p - q} f_{k -
q} - f_{p - q} f_p - f_{p - q}] \;. 
\end{eqnarray}

We have considered only the imaginary parts, since they are part of the collisional term in \eqref{collisional}. We ignored the real parts since they only will contribute to small shifts in the energies.

\subsection{Non-coherent Particle Keldysh Self-energy $\Sigma_{(nc)}^K$}

The contributions to the Keldysh self-energy $\Sigma_{(nc)}^K$ are given by
\begin{equation}
- i \Sigma_{(nc)}^K =  - i \Sigma_{(a)}^K - i \Sigma_{(b)}^K \;, 
\end{equation}
where as before, we have split them in two parts, defined as
\begin{eqnarray}
- i \Sigma_{(a)}^K &=& \frac{g^2}{4} \Phi_0(x')  \bigg[ 2 G_{(nc)}^K (x, x')^2 + 4
G_{(nc)}^A (x', x) G_{(nc)}^R (x, x') + 4 G_{(nc)}^R (x', x) G_{(nc)}^A ( x, x' ) \nonumber\\
& & + 4 G_{(nc)}^K (x, x') G_{(nc)}^K (x', x) + 2 G_{(nc)}^R (x, x')^2 + 2 G_{(nc)}^A (x, x')^2 \bigg]\Phi_0^\ast(x) \;, \label{sigmaaK}\\
- i \Sigma_{(b)}^K &=& i \frac{g^2}{4} \bigg[ 2 G_{(nc)}^K (x', x) G^R (x, x')^2 + 2
G_{(nc)}^K (x', x) G_{(nc)}^K (x, x')^2 \nonumber\\
& & + 4 G_{(nc)}^A (x', x) G_{(nc)}^K ( x, x' ) G_{(nc)}^R (x, x')
+ 2 G_{(nc)}^K (x', x) G_{(nc)}^A (x, x')^2 \nonumber\\
& & + 4 G_{(nc)}^R (x', x) G_{(nc)}^A (x, x') G_{(nc)}^K (x, x') \bigg] \;. \label{sigmabK}
\end{eqnarray}
We use the identities
\begin{eqnarray}
G^R (x', x) G^A (x, x') + G^A (x', x) G^R (x, x') &=& - (G^R (x', x) - G^A
(x', x)) (G^R (x, x') - G^A (x, x')) \;, \\
G^A (x, x')^2 + G^R (x, x')^2 &=& (G^R
(x, x') - G^A (x, x'))^2 \;, 
\end{eqnarray}
we symmetrize the last term in $\Sigma_{(b)}^K$ and following the same steps as in the previous case, we get that
\begin{eqnarray}
\Sigma_{(a)}^K &=& -2i\pi g^2  n_c \underset{\mathbf{p}}{\sum}\bigg[4 \delta (\varepsilon_c + \varepsilon_{\mathbf{p}} -
\varepsilon_{\mathbf{k}} - \varepsilon_{\mathbf{p} -
\mathbf{k}}) (2 f_{p - k} f_p + f_p + f_{p - k}) \nonumber\\
& & + 2 \delta
(\varepsilon_c + \varepsilon_{\mathbf{k}} -
\varepsilon_{\mathbf{p}} - \varepsilon_{\mathbf{k} -
\mathbf{p}}) (2 f_{k - p} f_p + f_p + f_{k - p} + 1)\bigg] \;, \\
\Sigma_{(b)}^K &=& -4i \pi g^2\underset{\mathbf{p},
\mathbf{q}}{\sum} \delta (\varepsilon_{\mathbf{p}} +
\varepsilon_{\mathbf{k} - \mathbf{q}} - \varepsilon_{\mathbf{k}} -
\varepsilon_{\mathbf{p} - \mathbf{q}}) [2 f_{p - q} f_{k - q} f_p + f_{p -
q} f_p + f_{p - q} f_{k - q} + f_{p - q} + f_{k - q} f_p] \;. 
\end{eqnarray}

\subsection{Collisional terms}

Using the expression in equation \eqref{collisional} we can generate two kind of collisional terms, constructed with the computed retarded and Keldysh self-energies: 
\begin{eqnarray}
I_{a}[f(k)] &=& i \Sigma_{(a)}^K+2(2f(k)+1)\Im(\Sigma_{(a)}^R) \;, \\
I_{b}[f(k)] &=& i \Sigma_{(b)}^K+2(2f(k)+1)\Im(\Sigma_{(b)}^R) \;. 
\end{eqnarray}
Explicitly, after some algebra and passing the sums to integrals, we can get that
\begin{eqnarray}
I_{a} &=& 4 g^2 n_c \int \frac{d^3 p_1 d^3 p_2 d^3 p_3}{(2 \pi)^2} \, \delta (\varepsilon_c +
\varepsilon_{\mathbf{p}_1} - \varepsilon_{\mathbf{p}_2} -
\varepsilon_{\mathbf{p}_3}) \, \delta (\mathbf{p}_1 - \mathbf{p}_2
- \mathbf{p}_3) \bigg( \delta (\mathbf{p}_1 - \mathbf{p}) - \delta (\mathbf{p}_2 -
\mathbf{p}) \nonumber\\ 
& & - \delta (\mathbf{p}_3 - \mathbf{p}) \bigg) 
\bigg[(1 + f_1) f_2 f_3 - f_1 (1 + f_2) (1 + f_3)\bigg] \;, \\
I_{b} &=& 4 g^2 \int \frac{d^3 p_2 d^3 p_3 d^3 p_4}{(2 \pi)^5}
\, \delta (\varepsilon_{\mathbf{p}_3} + \varepsilon_{\mathbf{p}_4} -
\varepsilon_{\mathbf{p}} - \varepsilon_{\mathbf{p}_2}) \, \delta (\mathbf{p}_3
+ \mathbf{p}_4 - \mathbf{p} - \mathbf{p}_2) \nonumber\\
& & \times \bigg[ f_3 f_4 (1 + f) (1 + f_2) -
f f_2 (1 + f_3) (1 + f_4) \bigg] \;. 
\end{eqnarray}

\bibliographystyle{unsrt}
\bibliography{Refs_cosmo}
 
\end{document}